# Dual-stage structural response to quenching charge order in magnetite

Wei Wang[1,*], Junjie Li[1], Lijun Wu[1], Jennifer Sears[1], Fuhao Ji[2], Xiaozhe Shen[2], Alex H. Reid[2], Jing Tao[1,#], Ian K. Robinson[1,3], Yimei Zhu[1,*], Mark P. M. Dean[1,*]

[1]Condensed Matter Physics and Materials Science Division, Brookhaven National Laboratory, Upton, NY 11973, USA
[2]SLAC National Accelerator Laboratory, Menlo Park, CA 94025, USA
[3]London Centre for Nanotechnology, University College, London WC1E 6BT, UK
[#]Current address: Department of physics, University of Science and Technology of China, 230026 Hefei, Anhui, China
Corresponding authors: wwang@bnl.gov; zhu@bnl.gov; mdean@bnl.gov

## Abstract

The Verwey transition in magnetite ($Fe_3O_4$) is the prototypical metal-insulator transition and has eluded a comprehensive explanation for decades. A major element of the challenge is the complex interplay between charge order and lattice distortions. Here we use ultrafast electron diffraction (UED) to disentangle the roles of charge order and lattice distortions by tracking the transient structural evolution after charge order is melted via ultrafast photoexcitation. A dual-stage response is observed in which $X_3$, $X_1$ and $\Delta_5$ type structural distortions occur on markedly different timescales of 0.7-3.2 ps and longer than 3.2 ps. We propose that these distinct timescales arise because $X_3$-type distortions strongly couple to the trimeron charge order whereas the $\Delta_5$-distortions are more strongly associated with monoclinic to cubic distortions of the overall lattice. Our work aids in clarifying the charge-lattice interplay using UED method and illustrates the disentanglement of the complex phases in magnetite.

## I. Introduction

Studying the mechanism of the metal-insulator transition in magnetite ($Fe_3O_4$), termed the Verwey transition, is one of the classic topics in condensed matter physics [1,2]. A key element of the transition was identified previously in structural studies as three-site small polarons named "trimerons" in which Fe sites experience a charge modulation of $Fe^{2.5+\delta}$-$Fe^{2.5-\delta}$-$Fe^{2.5+\delta}$, where $\delta$ quantifies the degree of charge modulation [3–6]. In addition to the charge order and metal-insulator transition [7–10], further changes in magnetization [11,12], orbital order [13–15], and overall crystal structure occur [16–19], which makes determining the Verwey transition mechanism in $Fe_3O_4$ challenging and fascinating.

Hitherto, experimental and theoretical studies have suggested that cooperative electron-phonon behavior is crucial for the Verwey transition [13,20–24]: the intimate coupling between electrons in Fe $t_{2g}$ orbitals and the phonon modes is thought to enhance electron localization in trimerons and effectively reduce the total energy of the system [25,26]. On the other hand, the stabilized charge ordered state modifies the interatomic interactions and may contribute to a concomitant structural instability [27,28].

Using conventional equilibrium methods, it is hard to explore the electron and lattice degrees of freedom separately due to the bi-directional interactions between them [29,30]. X-ray and optical pump-probe measurements have demonstrated photoinduced destruction of charge order and phase separation yielding metallic and insulating regions in the monoclinic-phase of $Fe_3O_4$ [31,32]. Compared with the pervious ultrafast studies on magnetite, we are more interested in the detailed lattice deformations related to the phonon modes in different and longer timescales. In this study, we use ultrashort laser pulses to decouple the electrons and lattice in the far-from-equilibrium state and employ MeV electron pulses to probe the charge ordered state and lattice deformation in the time domain [33,34]. Due to the energy redistribution between electrons and lattice in the photoexcited system, we observed energy flow from the electrons to the lattice. Taking advantage of MeV electron pulses with access to large regions of momentum space, we propose a pathway for the energy flow: the degree of charge order in the trimerons is firstly weakened in a short time delay (0-0.7 ps), which is consistent to the characteristic timescale of metallic and insulating phase separation mentioned in [31]. Then, we found a two-stage lattice response: due to electron-phonon interactions, the lattice distorts via dominant $X_3$- and $X_1$-type displacements from 0.7 ps to 3.2 ps, and lastly $\Delta_5$-type structural deformation emerges in the second stage after 3.2 ps in the phonon-phonon decay process. Our ultrafast electron diffraction (UED) experiment provides a picture of how the electron and lattice subsystems interact with each other induced by incident photons, which advances the fundamental understanding of the Verwey transition.

## II. Methods

### A. Sample preparation

Single crystal $Fe_3O_4$ was purchased from SurfaceNet GmbH, Germany. The resistance measurement result shows a first-order phase transition at 115 K, see Supplemental Material at [35] for the electronic resistance measurement result (see, also, reference [36] therein). The structure transition from high temperature cubic phase to low-temperature monoclinic phase has been confirmed using electron diffraction method, and the corresponding electron diffraction patterns captured from monoclinic phase and cubic phase along $[100]_{cubic}$ orientation are shown in Figs. S1b, 1c. The 115 K phase transition temperature suggests the possible presence of oxygen vacancies [9]. The UED sample was prepared by mechanical polishing, Focused Ion Beam (FIB) milling and is thinned until electron transparent. The sample orientation is along the $[100]_{cubic}$ orientation. The sample size is about 100 μm × 80 μm × 100 nm. Besides the FIB sample preparation, we also prepared the sample using another two methods: 1) mechanical polishing with Ar ion milling; 2) crushed sample suspended on the TEM Cu grid. The samples prepared using these three methods were tested in the *in situ* cooling TEM experiment. The structural phase transition was observed at ~ 115 K in all the samples, indicating that the FIB preparation method does not change the sample quality.

### B. MeV ultrafast electron diffraction

The ultrafast electron diffraction experiments were performed at the MeV UED beamline at SLAC National Accelerator Laboratory [37]. The 4.2 MeV electron pulses of < 150 fs (FWHM) duration transmitted through the sample at the normal incidence. The sample was excited using a 1.55 eV laser pulse with a duration of 75 fs (FWHM) at a repetition rate of 180 Hz. The pump fluences were 1, 2, 3, 5, 7.5 $mJ\cdot cm^{-2}$ in the experiment. The sample was cooled down to 34 K using a conducting sample holder during the experiment.

### C. Electron diffraction simulation

To understand the intensity variation of the experimentally measured superlattice (SL) reflections and Bragg peaks we carried out dynamic electron diffraction simulations considering charge distribution and lattice distortion in the system. The simulation is based on the Bloch wave method using the computer codes developed in-house. The dynamic and multiple scattering effect of the electrons has been considered in the simulation. The simulation results are consistent with the experimental observations, which cannot be explained using kinematic scattering theory. At 34 K experimental temperature, there are multiple twin variants appearing in the monoclinic phase, i.e., along $[110]_{\text{monoclinic}}$, $[1\bar{1}0]_{\text{monoclinic}}$ and $[001]_{\text{monoclinic}}$ directions, which are equivalent in the high-temperature cubic phase. In the electron diffraction simulation, all the twin variants were considered, i.e., the diffraction pattern along each orientation was simulated and then averaged all the patterns. Additionally, we considered other factors in the simulation, i.e., sample thickness and sample bending effect. In our diffraction simulation codes, we calculate the diffraction pattern for a specific sample thickness or a series of thickness. In this study, considering the large sample area, the sample thickness slightly changes from area to area. We simulated the diffraction pattern at a series of thickness ($t$), i.e., $t$ = 60 nm - 79 nm, $\Delta t$ = 1 nm. Namely, the diffraction pattern was simulated using twenty different thickness, e.g., 60 nm, 61 nm, …, 79 nm and the twenty diffraction patterns were averaged to form one diffraction pattern. When the sample is thinned down for electron beam transparent, the bending effect is not evitable. To simulate the sample bending effect, we tilt the electron beam in the simulation as a precession electron beam. The simulated electron diffraction patterns with and without considering the precession angles are shown in Fig. S2. Without considering the precession angles in the simulation, the intensity of the higher-order Bragg peaks decreases fast, which is not consistent with the experimental pattern. With the precession angles, the intensities of the Bragg peaks and SL reflections are close to the experimental data.

## III. Experiment and Results

Figure 1a shows the MeV UED setup for the study. The pump-probe electron diffraction experiment was performed using a 4.2 MeV electron probe with 1.55 eV laser pump pulses [37]. Above the Verwey transition at ~ 115 K, $Fe_3O_4$ has a cubic unit cell with $Fd\bar{3}m$ space group No. 227 and lattice constant $a \approx 8.4$ Å [16]. Unless otherwise stated, we index diffraction patterns using this unit cell. In Fig. 1a, we show the diffraction pattern at 34 K well below the Verwey transition, with the incident beam along the $[100]_{cubic}$ direction. The extinction rules for the cubic unit cell mean that this phase has Bragg peaks of the type $(0, k, l)_{cubic}$ where $k + l = 4n$ and $n$ is an integer [38]. These Bragg peaks are present above and below the Verwey transition. Below the transition, the sample undergoes an approximate $\sqrt{2} \times \sqrt{2} \times 2$ type reconstruction into a monoclinic structure with $Cc$ space group No. 9 with $a$ = 11.88 Å, $b$ =11.85 Å and $c$ =16.78 Å and $\alpha = \gamma = 90°$, $\beta = 90.236°$ [39]. During this transition, each of the principle cubic axes can transform into either the $[001]_{monoclinic}$, $[110]_{monoclinic}$, or $[1\bar{1}0]_{monoclinic}$ directions. This means that each of the Bragg peaks generates additional crystallographically equivalent spots due to the six twining-related monoclinic domains [40]. Additionally, $(h, k, l)_{cubic}$ SL peak satellites appear at positions, e.g., $(h, 0, 0)_{cubic}$, $h$ = 2, 6, etc. Furthermore, if the $[110]_{monoclinic}$ and $[1\bar{1}0]_{monoclinic}$ direction lies in the diffraction plane, additional half-integer SL peaks become visible, such as $(h, 0, l+1/2)_{cubic}$, and $(0, k, l+1/2)_{cubic}$. All these SL peaks have been previously assigned to trimeron order effects, where each trimeron consists of a linear unit of three Fe ions with a displacement of the two outer $Fe^{3+}$ ions towards the central $Fe^{2+}$ ion in the monoclinic phase [41,42]. A detailed illustration of how the monoclinic phase and twinning structure affect the pattern is provided in Appendix A.

**A. Intensity variation of SL reflections**

An enlarged view of four Bragg peaks $[(0, \bar{2}, 2), (0, \bar{4}, 4), (0, \bar{2}, 6), (0, 0, 4)]$ with surrounding SL reflections is shown as the inset to Fig. 1b. The time dependence of each SL peak has been measured. The average intensity variation of six SL reflections [ $\left(0, \bar{2}, 2 + \frac{1}{2}\right)$, $(0, \bar{2}, 3)$, $\left(0, \bar{2}, 3 + \frac{1}{2}\right)$, $\left(0, \bar{2}, 4 + \frac{1}{2}\right)$, $(0, \bar{2}, 5)$, $\left(0, \bar{2}, 5 + \frac{1}{2}\right)$] as a function of time delay after the laser pump with 5 $mJ \cdot cm^{-2}$ fluence is shown in Fig.1b. A fast drop is seen within the first 0.7 ps followed by slower decays at longer timescales. Each SL reflection intensity variation was measured individually, and the intensities from SL reflections with ($h$, $k$, $l$ = half-integers) and ($h$, $k$, $l$ = integers) were analyzed separately, which are related to charge orders with different periodicities [43]. We found that the SL intensity variations are similar for different sets of SL peaks in the reciprocal space. To increase the signal-noise ratio, an averaged intensity measurement result from 32 SL reflections is shown in Fig. S3. Apart from its intensity variation, the SL peak profile also changes, as shown in Fig. 1c, indicating that the SL peaks become broader with time (Fig. 1d). The measurements from the SL reflections illustrate that the peak intensities get weaker and the correlation length of the trimeron order in real space becomes shorter after photoexcitation.

To more fully understand the nature of the photo-induced sample changes, we also examined the time-dependent response from the Bragg peaks. Figure 2a shows the intensity difference maps for a selection of time delays, which display the overall temporal evolution of the Bragg peaks (After ~20 ps, only small changes in the distribution of the intensity variations were observed). Here, the intensity difference map is $\Delta I(\boldsymbol{q}, t) = I(\boldsymbol{q}, t) - I(\boldsymbol{q}, t_0)$, where $t_0$ represents the time before the pump arrives and $\boldsymbol{q}$ is the scattering vector in reciprocal space. During the early timescale (0-0.7 ps), the intensities of Bragg peaks increase slightly, which is the opposite behavior compared to the SL reflections. The SL intensity is reduced by ~ 80%, while the Bragg peak intensity only increases by ~ 2% in the first 0.7 ps. As is known from equilibrium studies [6], the SL reflections in $Fe_3O_4$ are more sensitive to the change of charge ordering than the Bragg peaks. This is confirmed by our electron diffraction simulations in which we predict how the diffraction pattern changes with and without charge order. A reduced charge discrepancy of the Fe ions was shown to reduce the intensity of SL reflections and increases the Bragg peak intensity in the first 0.7 ps (see Appendix B). The result demonstrates that charge order in trimerons gets quenched in the first 0.7 ps.

**B. Intensity variations of Bragg peaks**

Figures 2b-2e plot representative intensity evolutions measured from four Bragg peaks, which clearly demonstrate different dynamic behaviors in different time regimes. After 0.7 ps, all the intensities start to decrease. Then, the intensity change becomes $\boldsymbol{q}$-dependent after ~ 3.2 ps, which results in the intensity difference distribution at 9.2 ps shown in Fig. 2a. The intensity from the Bragg peaks with low-index $\boldsymbol{q}$, e.g., (0, 4, 0), is lower than that at time zero $t_0$, and the intensity with high-index $\boldsymbol{q}$, e.g., (0, 8, 0), is higher than that at $t_0$. After 9.2 ps, the intensity of the reflections near the central horizontal direction, e.g., (0, 0, 8), starts to decrease, leading to a negative intensity as shown at 19.2 ps. Since the experimental temperature is 34 K, the sample is in the monoclinic phase, which breaks the cubic symmetry, leading to the different temporal behavior between (0, 4*n*, 0) and (0, 0, 4*n*) reflections. Additionally, in the low-temperature phase, the SL formation originates from the charge order and lattice distortion, hence the change from both of them affects the SL intensity. After the fast intensity drop induced by the charge order quenching, the SL intensity change becomes slow after 0.7 ps, the intensity of Bragg peaks starts to exhibit obvious changes at the same time, indicating lattice distortions are triggered following the charge order quenching, through electron-phonon and phonon-phonon interactions in the strongly correlated materials [44–46]. Since the Bragg peaks captured from large range of the momentum space are more sensitive to the lattice structure, comparing to the SL reflections, we are going to figure out the lattice dynamic behavior using the intensity variations of Bragg peaks.

Due to the complex structural transformation from the high-temperature cubic phase to low-temperature monoclinic phase, several phonon modes have been proposed to be major factors in the phase transition. The phonon modes of the cubic structure at the Δ point and X point are most popular and highly debated [25–27,42,47]. Neutron scattering experiments observed a strong reflection at (*h*, 0, *l*+1/2) reciprocal points, which indicates that the phonon modes with the wave vector $\boldsymbol{k}$ = (0, 0, 1/2) at the Δ point become unstable at the phase transition and lead to a doubled lattice parameter along the $\boldsymbol{c}$ axis. $\Delta_4$ and $\Delta_5$ phonon modes are mainly discussed in [48], and it was found that both of these contribute to the SL reflections with half-integers in the low-temperature phase. However, the atomic displacements with $\Delta_4$ and $\Delta_5$ modes cannot fully explain the weak critical scattering at the X points of the Brillouin zone at reflections $(0, k, l)_{\text{cubic}}$, $k+l = 4n+2$ (*n* is an integer). Moreover, X-ray and neutron studies claimed that $X_1$ and $X_3$ phonon modes with $\boldsymbol{k}$ = (0, 0, 1) contribute to the phase transition as well [31,39,49,50]. The W modes with $\boldsymbol{k}$ =

(1/2, 1, 0) play an important role in the formation of modulated structures in the low-temperature phase, and the contribution of $X_4$ modes improves the structural refinement [51].

Based on the above considerations, we construct atomic models with atomic displacements based on the $W_1$, $W_2$, $\Delta_4$, $\Delta_5$, $X_1$, $X_3$, $X_4$-type phonon modes [26,39,51]. The displacement direction in each atomic layer and the relative displacement amplitudes and phases in each phonon mode are based on [26,39,48,50,52,53], where the patterns of lattice distortion in the primitive unit cell for these seven phonon modes are listed. The atoms in $Fe_3O_4$ can be categorized into three types: Fe atoms on the tetrahedral site, Fe atoms on the octahedral site and oxygen atoms. To figure out the intensity changes induced by these atomic species, we simulated the diffraction patterns considering the displacements of Fe atoms in the tetrahedra, Fe atoms in the octahedra and the oxygens, separately. We found that the intensity change induced by the displacement of the Fe atoms in the octahedra is about four times larger than that induced by the Fe atoms in the tetrahedra and the oxygen atoms using the same displacement amplitude in the same type of lattice distortion. Additionally, the Fe atoms on the octahedral sites are directly related to the electronic order, i.e., trimeron lattice, in magnetite. Therefore, the change of the charge order induced by the incident photon mainly affects the octahedral Fe atoms through the electron-lattice interaction [31]. To study the electron-lattice interplay in magnetite, we made a crude approximation and mainly focused on the atomic displacements of Fe atoms on the octahedral sites, i.e., atoms in trimerons, which is the same way as shown in [31]. Furthermore, since the UED sample is in the low-temperature monoclinic phase before time zero, in the atomic displacement models, we transformed and expanded the atomic displacements into the monoclinic system as illustrated in Fig 3a. In the monoclinic phase, the principal axes $\boldsymbol{x}$ and $\boldsymbol{y}$ are rotated by ~45°, the lattice parameter along $\boldsymbol{z}$ is doubled. The detailed coordinates of Fe atoms with the atomic displacements used in the diffraction simulation are listed in Tables I-VII in Appendix B.

We calculated how the electron diffraction pattern varies as a function of amplitude of the $W_1$, $W_2$, $\Delta_4$, $\Delta_5$, $X_1$, $X_3$, $X_4$-type lattice distortions, finding that each type of distortion produces a distinct modification of the reflection intensities in the pattern. To make the diffraction simulation close to the experimental conditions, we considered multiple factors. According to the SL reflections distributed in the UED pattern, we infer that the experimental monoclinic crystal consists of $[001]_{monoclinic}$ domain and its 90°-rotation domain, $[110]_{monoclinic}$ and $[1\bar{1}0]_{monoclinic}$ domain, i.e., a total of four domain structures. The simulated diffraction patterns from the multiple domains are averaged. Additionally, the sample thickness variation and sample bending are considered in the simulations by averaging a series of diffraction patterns with thicknesses and crystal orientations in certain ranges, 60-79 nm in thickness and 0.5°, 0.8°, 1.5°, 1.8° precession angles in crystal orientation, obtained by fitting the experimental diffraction pattern taken before photoexcitation.

By comparing the calculated results for the predicted models with the data in Fig. 2a, it was possible to elucidate the lattice distortion trajectories as illustrated in Fig. 3a. The quenched charge ordered state drives the atomic displacement following the $X_3$ phonon modes after 0.7 ps. The intensity variation ($\Delta I$) caused by the changes from electronic structure and lattice structure is shown in the simulated result in Fig. 3b. As we mentioned in Section A, the quenched charge order decreases the SL intensity and increases the Bragg peak intensity. Due to the lattice distortion following $X_3$ phonon modes, the intensity of Bragg peaks starts to decrease at 0.7 ps, as shown in Figs. 2b-2e. After ~ 3.2 ps, the atomic displacement follows a transverse acoustic (TA) $\Delta_5$ phonon, which we term $\Delta_5$ *mode-x, y*, indicating the atomic displacement is predominantly in the ***x-y*** plane. The displacement of the octahedral-site Fe atoms in one ***x-y*** plane layer moves in the same direction with a constant deviation, however, the relative displacement amplitude among the layers follows a sinusoidal variation along the ***z*** direction (see Fig. 3a). Since the wave vector ***k*** of the $\Delta_5$ phonon mode is (0, 0, 1/2), the displacement direction in the first four layers is along the -***x***+***y*** axis and is along the ***x-y*** axis in the next four layers, as shown in Fig. 3a. Since $\Delta_5$ is a two-dimensional representation, another set of symmetry-related displacements are present along -***x-y*** and ***x***+***y*** directions. These two models provide similar results in the diffraction simulation shown in Fig. 3c. After 3.2 ps, the intensities of (0, 4, 0) and (0, 0, 4) reflections continue to decrease, and the intensities of (0, 8, 0) and (0, 0, 8) reflections start to increase due to the lattice deformation following $\Delta_5$ *mode-x, y*, which is consistent with the experimental observations shown in Figs. 2b-2e. After 9.2 ps, the atomic displacements along the ***z*** direction increase following the $\Delta_5$ phonon mode pattern, i.e., a $\Delta_5$ *mode-x, y, z*, emerges, leading to the intensity decrease at (0, 0, $l$), $l = 4n$, $n$ is an integer. For Bragg peaks at (0, $k$, 0), $k = 4n$, the displacement along ***z*** direction slightly increases the intensity at (0, 4, 0) and decreases the intensity at (0, 8, 0) from 9.2 ps to 19.2 ps. The corresponding simulation result is in Fig. 3d. Compared with the lattice distortion in $\Delta_5$ modes, the atomic displacement amplitudes are the same in each layer in $\Delta_4$ modes as shown in Table V. The simulated results based on $\Delta_4$ modes shown in Fig. S4a are not consistent with the experimental data shown in Fig. 2a.

According to the electron diffraction simulation for $X_1$-, $X_3$- and $X_4$-type displacements, we found that the atomic displacements following $X_1$ and $X_3$ phonon modes give similar impacts on the intensity of Bragg peaks, i.e., both modes reduce the intensity. $X_1$ and $X_3$ phonon modes have the same wave vector ***k*** = (0, 0, 1) in reciprocal space, but the atomic displacement in the Fe-O layer is different. In the pattern of atomic displacements with $X_1$ mode, the two neighboring Fe atoms in the same Fe-O plane move in opposite directions, e.g., along ***x***+***y*** axis and -***x-y*** axis, respectively. In the case of the $X_3$ mode, the atomic displacement of the Fe atoms in one layer is along the same direction as shown in Fig. 3a. The displacement amplitude in each layer is the same, and the displacement periodicity along the ***z*** direction coincides with the lattice constant in the cubic phase, i.e., half the lattice constant in the monoclinic phase. In the simulation model with the $X_3$ mode in Fig. 3a, the displacement direction from the first layer to the fourth layer is along -***x***+***y***, -***x-y***, ***x-y,*** and ***x***+***y*** axis, respectively, and repeats in the next four Fe-O layers. The lattice distortions following $X_1$ and $X_3$ phonon modes display similar results to the intensity variation of the Bragg peaks. In $X_4$ phonon modes, the lattice distortion is similar to $X_3$ modes, but the atomic displacement amplitude changes from layer to layer along ***z*** direction shown in Table III. The simulated diffraction result (in Fig. S4b) with $X_4$-type displacements is not fully consistent with the experimental result as shown in Fig. 2a. According to the simulation, the lattice distortion with the $X_1$ phonon mode cannot be larger than 0.023 Å, since a further displacement makes the intensity variation inconsistent with the experimental result. However, comparing with the $X_1$ mode, the displacement along the $X_3$ mode works well with relatively larger deviations $\delta$ up to 0.060 Å. Therefore, we conclude that the lattice distortions following the phonon modes with $X_1$ and $X_3$ symmetries dominate from 0.7 ps to 3.2 ps.

What's more, the intensity variation induced by the lattice distortions following $W_1$ and $W_2$ phonon modes were calculated shown in Fig. S4c, which are incompatible with the experimental intensity variations. According to the series simulation result, we conclude that $X_1$, $X_3$ and $\Delta_5$ phonon modes is a solution that is physically reasonable and consistent with the data, which are a most likely scenario. We cannot systematically exclude every conceivable motion of all the atoms in the charge-ordered unit cell.

## IV. Discussions

### A. Intensity vs. pump fluence

The pump fluence dependence of the effects were studied at 1, 2, 3, 5, 7.5 $mJ \cdot cm^{-2}$. The SL reflections (Fig. 4a) all show a similar response: the intensity drops over a short time delay then becomes relatively flat at low pump fluences, 1 and 2 $mJ \cdot cm^{-2}$. With higher pump fluences, the intensity slowly changes after 0.7 ps, and the intensity becomes flat until ~ 50 ps at 5 and 7.5 $mJ \cdot cm^{-2}$. The intensity variation value $|\Delta I|$ at 9.2 ps at different fluences is extracted and shown in Fig. 4b. The plot manifests that at low fluences, the intensity variation is proportional to the fluence. When the fluence reaches 7.5 $mJ \cdot cm^{-2}$, the SL intensity change is slightly larger than that at 5 $mJ \cdot cm^{-2}$, which suggests the charge ordering phase get less sensitive to the incident photon and starts to saturate above a pump fluence value of about 5 $mJ \cdot cm^{-2}$.

The Bragg peaks respond differently. Our measurement results shows that the Bragg peak intensity exhibits minor changes at 1 $mJ \cdot cm^{-2}$ and 2 $mJ \cdot cm^{-2}$ pump fluences. Above 3 $mJ \cdot cm^{-2}$, we start to observe the intensity changes of Bragg peaks. The intensity variation $|\Delta I|$ at 3 $mJ \cdot cm^{-2}$ is ~ 1/2 of that at 5 $mJ \cdot cm^{-2}$, and the $\Delta I$ at both 3 and 5 $mJ \cdot cm^{-2}$ show a similar tendency in first 9.2 ps, which implies the $X_1$, $X_3$ and $\Delta_5$ phonon modes are involved in the dynamic process at 3 $mJ \cdot cm^{-2}$ as well. The intensity difference maps at 3 $mJ \cdot cm^{-2}$ fluence are shown in Fig. S5. However, the intensity from the Bragg reflections, highlighted by the frame in Fig. 4c, does not decrease at long-time delays, which is different from the intensity change at 5 $mJ \cdot cm^{-2}$. Figures 4c and 4d show the intensity difference map taken at ~ 60 ps with the pump fluence of 3 $mJ \cdot cm^{-2}$ and 5 $mJ \cdot cm^{-2}$, respectively. According to the intensity variation and lattice distortion after 9.2 ps at 5 $mJ \cdot cm^{-2}$, we infer that there is no obvious atomic movement along the ***z*** direction in the case with the pump fluence of 3 $mJ \cdot cm^{-2}$. The simulation for the long-time delay is shown in Fig. S5h. We note that the intensity variation at 7.5 $mJ \cdot cm^{-2}$ is similar to that at 5 $mJ \cdot cm^{-2}$ (Fig. S6), and the SL intensity shows a subtle difference between these two pump fluences. We thus can conclude that the dynamic pathway at 7.5 mJ·cm$^{-2}$ is similar to that at 5 mJ·cm$^{-2}$, i.e., $X_1$, $X_3$ and $\Delta_5$ modes appear at distinctive timescales.

The pump fluence dependent observation reveals that at relatively low pump fluence, the excited electrons are not sufficient to drive the lattice distortion corresponding to the phonon modes. Above the threshold pump fluence (~ 3 mJ·cm$^{-2}$), the atom movement follows different phonon modes at different timescales. At relatively high pump fluences, approaching 7.5 mJ·cm$^{-2}$, the structural distortion becomes almost independent of fluence, since the charge ordering state is almost completely quenched. In addition, the atomic displacement with $\Delta_5$ phonon modes along the $\boldsymbol{z}$ direction occurs later than the occurrence of the displacement in the $\boldsymbol{x}$ and $\boldsymbol{y}$ directions, indicating an anisotropic phonon dispersion along <001> in the monoclinic phase. Moreover, inelastic neutron scattering studies reported that the dispersion of the Γ-Δ-X TA modes with polarization along the [001] is slightly harder than for the displacement vector aligned to [100]; this is associated with the differences of the averaged charge density modulation in the $\boldsymbol{x}$-$\boldsymbol{y}$ plane and along the $\boldsymbol{z}$ direction [47]. These findings imply that the atomic displacement along the $\boldsymbol{z}$ direction require more energy than the displacement in the $\boldsymbol{x}$ and $\boldsymbol{y}$ directions in the monoclinic phase, which could be a reason for the absence of the $\boldsymbol{z}$-displacement at the low pump fluence of 3 mJ·cm$^{-2}$.

**B. Two timescales for the lattice distortions related to the phonon modes**

The photoinduced dynamic process has been summarized in Fig. 5, including electron excitation and lattice distortions. In the structural deformation following the charge-order quenched state, we have identified two distinctive timescales: I) The atomic displacements following the $X_3$ and $X_1$ optical phonon modes get excited right after the melting of the charge ordering state at 0.7 ps, which demonstrates an energy flow from electrons to phonons. The lattice distortions with the corresponding X symmetry are preeminent from 0.7 ps to 3.2 ps in Stage I; II) After 3.2 ps, the lattice distortion with the $\Delta_5$ TA phonon modes enters Stage II, which could be stimulated via phonon-phonon interactions in the relaxation process.

In group theory studies, phonon modes with $X_3$ and $\Delta_5$ symmetry have been identified as primary order parameters (OPs) for the structural transition from the high-temperature cubic structure to low-temperature monoclinic structure [25,26]. Furthermore, the $X_1$ phonon mode is a secondary OP, which couples to the first OP [54]. Thus, the excited $X_3$ phonon mode induced by the incident photons prompts the excitation of $X_1$ phonon mode. The timescales in Stage I for the excitations of $X_1$ and $X_3$ phonon modes are not distinguishable in the current data. As another primary OP for the phase transition, the $\Delta_5$ phonon mode with $\boldsymbol{k}$ = (0, 0, 1/2), is crucial for the doubling of the unit cell, which contributes more to the structure symmetry breaking compared with the $X_3$ phonon mode; its coupling strength to the electronic structure is relatively weak. The lattice distortion with $\Delta_5$ phonon modes is initiated by phonon-phonon coupling. Hence, the $\Delta_5$ phonon mode is observed at the second stage in the long timescale.

## V. Conclusion

In summary, we have isolated the electronic structure and lattice structure response of $Fe_3O_4$ in the time domain following ultrashort laser pulses of 800 nm wavelength and characterized their interactions by pump-probe electron diffraction. We observed the amplitude of the electronic ordering in the trimerons is significantly reduced within 0.7 ps after photoexcitation, demonstrating a quenching process of the charge order in the low-symmetry phase. Subsequently, a two-stage lattice response related to the phonon modes with different symmetries was observed. The emergence of the $X_3$ and $X_1$ phonon modes and the $\Delta_5$ TA phonon modes on different timescales substantiates their specific roles in the Verwey transition. The temporal evolution in the photoinduced system demonstrates the complex interplay between the charge and lattice degrees of freedom in magnetite and provides a deeper understanding of Verwey transition. The observations of femtosecond-timescale electronic structural excitation and picosecond-timescale lattice structural response to the induced photon are consistent with the x-ray-based observations in [31]. Additionally, the resonant x-ray scattering method has the capability to detect 2*p*-3*d* core valence resonance of Fe atoms, which is more sensitive to the electronic dynamic behaviors. The electron scattering method is able to access large range of momentum space, capturing more information about the crystal structure. The combination of these two methods allows us to reach a deeper and complete understanding of the interplay among these degrees of freedom in magnetite.

**Acknowledgements**
Work at BNL was supported by US Department of Energy (DOE), Office of Science, Office of Basic Energy Sciences (BES), Materials Sciences and Engineering Division under Contract No. DE-SC0012704. The experimental MeV-UED part of this research was performed at the SLAC MeV-UED facility, which is supported by the U.S. Department of Energy BES SUF Division Accelerator and Detector R&D program, the LCLS Facility, and SLAC under Award No. DEAC02-76SF00515. The UED samples were thinned using Focused Ion Beam at the Center for Function Nanomaterials at BNL, which is supported by DOE BES Users Facility Division. Work performed at UCL was supported by EPSRC. 

## APPENDIX A: DIFFRACTION PATTERNS IN THE HIGH-TEMPERATURE AND LOW-TEMPERATURE PHASES

During the structural phase transition in $Fe_3O_4$, the <001>$_{cubic}$ direction becomes six monoclinic domains in the low-temperature phase, along $[001]_{monoclinic}$, $[110]_{monoclinic}$, and $[1\bar{1}0]_{monoclinic}$ directions. Three of them are shown in Fig. 6b, another three domains are rotated 90° along $[001]_{monoclinic}$, $[110]_{monoclinic}$ and $[1\bar{1}0]_{monoclinic}$ directions. The different types of SL reflections in these domain structures are illustrated in Fig. 6c.

## APPENDIX B: ELECTRON DIFFRACTION SIMULATION

Figure 7a is the simulated result for the quenching of the charge ordering phase by reducing the charge discrepancy between the Fe ions in the trimerons. The intensity of SL reflections is highly reduced, and the intensity of the Bragg peaks is increased, which is consistent with the experimental observation at 0.7 ps. In the simulation, the charge ordering arrangement is based on the trimeron model in [6]. The valence charge states of Fe ions on the octahedral site are divided into : $Fe^{(2.5-\delta)+}$ and $Fe^{(2.5+\delta)+}$, which is shown in the Table S7 in [6]. Since there are 16 Wyckoff positions for the octahedral Fe ions, eight Fe ions' charge state is $Fe^{(2.5-\delta)+}$ and another eight Fe ions' charge state is $Fe^{(2.5+\delta)+}$. Using the in-house developed code, the charge state of Fe ions can be changed into fractional value, e.g., $Fe^{2.3+}$ and $Fe^{2.7+}$. Then the corresponding form factors will be recalculated. For example, the electron configuration of $Fe^{0+}$ is $3d^6$, and we change it into $3d^{3.7}$ for $Fe^{2.3+}$. To reduce the charge discrepancy, e.g., we changed the valence states in Fe ion from $Fe^{2.3+}$ and $Fe^{2.7+}$ to $Fe^{2.4+}$ and $Fe^{2.6+}$, and the atom form factors are recalculated for the new valence state.

We simulated one electron diffraction pattern for the charge ordering states $Fe^{2.3+}$ and $Fe^{2.7+}$ as an initial pattern ($P_0$) at time zero and simulated another diffraction pattern ($P_1$) for the charge ordering states $Fe^{2.4+}$ and $Fe^{2.6+}$ using the same simulation parameters. This pattern is the same simplified charge distribution model as that used by [6]. Then we did subtraction of the initial pattern ($P_0$) from pattern ($P_1$) and we got the different map, which is shown in Fig. 7a. The SL reflection intensity is reduced, and the Bragg peak intensity is increased, which is qualitatively consistent with the experiment data shown in Fig. 2a.

Based on the charge order quenching, we moved the atoms off the original positions. The displacement corresponds to $X_3$ and $X_1$ phonon modes in $Fe_3O_4$ presents a similar result. The atomic displacement pathway following $X_1$ and $X_3$ phonon modes are shown in Figs. 7b and 7c, respectively. The simulated intensity difference pattern based on the $X_3$ phonon mode is in Fig. 7d. The intensity distribution is similar to the result shown in Fig. 7a, but the difference of the Bragg peak intensity between Figs. 7a and 7d is shown in Fig. 7e, indicating the Bragg peak intensity is reduced as a result of the atomic displacement following the $X_3$ phonon mode.

The atomic displacement parameters of Fe atoms on the octahedral sites in $X_3$, $X_1$, $\Delta_5$, and $\Delta_4$-type lattice distortions are summarized in Tables I-IV. There are 64 Fe atoms on the octahedral sites in the monoclinic phase. The original atom coordinates in the monoclinic phase are listed in the tables.

**Table I** $X_3$-type atomic displacement. The Fe atomic coordinates are listed. δ is the magnitude of atomic displacement. $\delta_{max}$ = 0.005 and the corresponding displacement is 0.060 Å.

| Fe | x | y | z | Fe | x | y | z |
|---|---|---|---|---|---|---|---|
| 1 | 0.75057-δ | 0.99788 | 0.00227 | 33 | 0.87694 | 0.87945-δ | 0.37981 |
| 2 | 0.25057-δ | 0.00212 | 0.50227 | 34 | 0.37694 | 0.12055-δ | 0.87981 |
| 3 | 0.25057-δ | 0.49788 | 0.00227 | 35 | 0.37694 | 0.37945-δ | 0.37981 |
| 4 | 0.75057-δ | 0.50212 | 0.50212 | 36 | 0.87694 | 0.62055-δ | 0.87981 |
| 5 | 0.75116-δ | 0.49865 | 0.00111 | 37 | 0.87644 | 0.38747-δ | 0.38075 |
| 6 | 0.25116-δ | 0.50135 | 0.50111 | 38 | 0.37644 | 0.61253-δ | 0.88075 |
| 7 | 0.25116-δ | 0.99865 | 0.00111 | 39 | 0.37644 | 0.88747-δ | 0.38075 |
| 8 | 0.75116-δ | 0.00135 | 0.50111 | 40 | 0.87644 | 0.11253-δ | 0.88075 |
| 9 | 0.00187-δ | 0.5005 | 0.5017 | 41 | 0.62663 | 0.88662+δ | 0.12178 |
| 10 | 0.50187-δ | 0.4995 | 0.0017 | 42 | 0.12663 | 0.11338+δ | 0.62178 |
| 11 | 0.50187-δ | 5E-4 | 0.5017 | 43 | 0.12663 | 0.38662+δ | 0.12178 |
| 12 | 0.00187-δ | 0.9995 | 0.0017 | 44 | 0.62663 | 0.61338+δ | 0.62178 |
| 13 | 0.99743-δ | 7.6E-4 | 0.49693 | 45 | 0.62878 | 0.37462+δ | 0.12311 |
| 14 | 0.49743-δ | 0.99924 | 0.99693 | 46 | 0.12878 | 0.62538+δ | 0.62311 |
| 15 | 0.49743-δ | 0.50076 | 0.49693 | 47 | 0.12878 | 0.87462+δ | 0.12311 |
| 16 | 0.99743-δ | 0.49924 | 0.99693 | 48 | 0.62878 | 0.12538+δ | 0.62311 |
| 17 | 0.74758+δ | 0.75639 | 0.2526 | 49 | 0.87599 | 0.62482-δ | 0.37671 |
| 18 | 0.24758+δ | 0.24361 | 0.7526 | 50 | 0.37599 | 0.37518-δ | 0.87671 |
| 19 | 0.24758+δ | 0.25639 | 0.2526 | 51 | 0.37599 | 0.12482-δ | 0.37671 |
| 20 | 0.74758+δ | 0.74361 | 0.7526 | 52 | 0.87599 | 0.87518-δ | 0.87671 |
| 21 | 0.75929+δ | 0.2522 | 0.25367 | 53 | 0.87543 | 0.13087-δ | 0.37437 |
| 22 | 0.25929+δ | 0.7478 | 0.75367 | 54 | 0.37543 | 0.86913-δ | 0.87437 |
| 23 | 0.25929+δ | 0.7522 | 0.25367 | 55 | 0.37543 | 0.63087-δ | 0.37437 |
| 24 | 0.75929+δ | 0.2478 | 0.75367 | 56 | 0.87543 | 0.36913-δ | 0.87437 |
| 25 | 0.00255+δ | 0.74372 | 0.75188 | 57 | 0.62566 | 0.62776+δ | 0.12584 |
| 26 | 0.50255+δ | 0.25628 | 0.25188 | 58 | 0.12566 | 0.37224+δ | 0.62584 |
| 27 | 0.50255+δ | 0.24372 | 0.75188 | 59 | 0.12566 | 0.12776+δ | 0.12584 |
| 28 | 0.00255+δ | 0.75628 | 0.25188 | 60 | 0.62566 | 0.87224+δ | 0.62584 |
| 29 | 0.00214+δ | 0.24588 | 0.75191 | 61 | 0.62788 | 0.12601+δ | 0.12652 |
| 30 | 0.50214+δ | 0.75412 | 0.25191 | 62 | 0.12788 | 0.87399+δ | 0.62652 |
| 31 | 0.50214+δ | 0.74588 | 0.75191 | 63 | 0.12788 | 0.62601+δ | 0.12652 |
| 32 | 0.00214+δ | 0.25412 | 0.25191 | 64 | 0.62788 | 0.37399+δ | 0.62652 |

**Table II** $X_1$-type atomic displacement. $\delta_{max}$ = 0.002 and the corresponding displacement is 0.023 Å.

| Fe | x | y | z | Fe | x | y | z |
|---|---|---|---|---|---|---|---|
| 1 | 0.75057-δ | 0.99788 | 0.00227 | 33 | 0.87694 | 0.87945-δ | 0.37981 |
| 2 | 0.25057-δ | 0.00212 | 0.50227 | 34 | 0.37694 | 0.12055+δ | 0.87981 |
| 3 | 0.25057-δ | 0.49788 | 0.00227 | 35 | 0.37694 | 0.37945-δ | 0.37981 |
| 4 | 0.75057-δ | 0.50212 | 0.50212 | 36 | 0.87694 | 0.62055+δ | 0.87981 |
| 5 | 0.75116-δ | 0.49865 | 0.00111 | 37 | 0.87644 | 0.38747-δ | 0.38075 |
| 6 | 0.25116-δ | 0.50135 | 0.50111 | 38 | 0.37644 | 0.61253+δ | 0.88075 |
| 7 | 0.25116-δ | 0.99865 | 0.00111 | 39 | 0.37644 | 0.88747-δ | 0.38075 |
| 8 | 0.75116-δ | 0.00135 | 0.50111 | 40 | 0.87644 | 0.11253+δ | 0.88075 |
| 9 | 0.00187+δ | 0.5005 | 0.5017 | 41 | 0.62663 | 0.88662-δ | 0.12178 |
| 10 | 0.50187+δ | 0.4995 | 0.0017 | 42 | 0.12663 | 0.11338+δ | 0.62178 |
| 11 | 0.50187+δ | 5E-4 | 0.5017 | 43 | 0.12663 | 0.38662-δ | 0.12178 |
| 12 | 0.00187+δ | 0.9995 | 0.0017 | 44 | 0.62663 | 0.61338+δ | 0.62178 |
| 13 | 0.99743+δ | 7.6E-4 | 0.49693 | 45 | 0.62878 | 0.37462-δ | 0.12311 |
| 14 | 0.49743+δ | 0.99924 | 0.99693 | 46 | 0.12878 | 0.62538+δ | 0.62311 |
| 15 | 0.49743+δ | 0.50076 | 0.49693 | 47 | 0.12878 | 0.87462-δ | 0.12311 |
| 16 | 0.99743+δ | 0.49924 | 0.99693 | 48 | 0.62878 | 0.12538+δ | 0.62311 |
| 17 | 0.74758-δ | 0.75639 | 0.2526 | 49 | 0.87599 | 0.62482+δ | 0.37671 |
| 18 | 0.24758-δ | 0.24361 | 0.7526 | 50 | 0.37599 | 0.37518-δ | 0.87671 |
| 19 | 0.24758-δ | 0.25639 | 0.2526 | 51 | 0.37599 | 0.12482+δ | 0.37671 |
| 20 | 0.74758-δ | 0.74361 | 0.7526 | 52 | 0.87599 | 0.87518-δ | 0.87671 |
| 21 | 0.75929-δ | 0.2522 | 0.25367 | 53 | 0.87543 | 0.13087+δ | 0.37437 |
| 22 | 0.25929-δ | 0.7478 | 0.75367 | 54 | 0.37543 | 0.86913-δ | 0.87437 |
| 23 | 0.25929-δ | 0.7522 | 0.25367 | 55 | 0.37543 | 0.63087+δ | 0.37437 |
| 24 | 0.75929-δ | 0.2478 | 0.75367 | 56 | 0.87543 | 0.36913-δ | 0.87437 |
| 25 | 0.00255+δ | 0.74372 | 0.75188 | 57 | 0.62566 | 0.62776+δ | 0.12584 |
| 26 | 0.50255+δ | 0.25628 | 0.25188 | 58 | 0.12566 | 0.37224-δ | 0.62584 |
| 27 | 0.50255+δ | 0.24372 | 0.75188 | 59 | 0.12566 | 0.12776+δ | 0.12584 |
| 28 | 0.00255+δ | 0.75628 | 0.25188 | 60 | 0.62566 | 0.87224-δ | 0.62584 |
| 29 | 0.00214+δ | 0.24588 | 0.75191 | 61 | 0.62788 | 0.12601+δ | 0.12652 |
| 30 | 0.50214+δ | 0.75412 | 0.25191 | 62 | 0.12788 | 0.87399-δ | 0.62652 |
| 31 | 0.50214+δ | 0.74588 | 0.75191 | 63 | 0.12788 | 0.62601+δ | 0.12652 |
| 32 | 0.00214+δ | 0.25412 | 0.25191 | 64 | 0.62788 | 0.37399-δ | 0.62652 |

**Table III** $X_4$-type atomic displacement.

| Fe | x | y | z | Fe | x | y | z |
|---|---|---|---|---|---|---|---|
| 1 | 0.75057+δ | 0.99788 | 0.00227 | 33 | 0.87694 | 0.87945+0.3*δ | 0.37981 |
| 2 | 0.25057+δ | 0.00212 | 0.50227 | 34 | 0.37694 | 0.12055+0.3*δ | 0.87981 |
| 3 | 0.25057+δ | 0.49788 | 0.00227 | 35 | 0.37694 | 0.37945+0.3*δ | 0.37981 |
| 4 | 0.75057+δ | 0.50212 | 0.50212 | 36 | 0.87694 | 0.62055+0.3*δ | 0.87981 |
| 5 | 0.75116+δ | 0.49865 | 0.00111 | 37 | 0.87644 | 0.38747+0.3*δ | 0.38075 |
| 6 | 0.25116+δ | 0.50135 | 0.50111 | 38 | 0.37644 | 0.61253+0.3*δ | 0.88075 |
| 7 | 0.25116+δ | 0.99865 | 0.00111 | 39 | 0.37644 | 0.88747+0.3*δ | 0.38075 |
| 8 | 0.75116+δ | 0.00135 | 0.50111 | 40 | 0.87644 | 0.11253+0.3*δ | 0.88075 |
| 9 | 0.00187+δ | 0.5005 | 0.5017 | 41 | 0.62663 | 0.88662-δ | 0.12178 |
| 10 | 0.50187+δ | 0.4995 | 0.0017 | 42 | 0.12663 | 0.11338-δ | 0.62178 |
| 11 | 0.50187+δ | 5E-4 | 0.5017 | 43 | 0.12663 | 0.38662-δ | 0.12178 |
| 12 | 0.00187+δ | 0.9995 | 0.0017 | 44 | 0.62663 | 0.61338-δ | 0.62178 |
| 13 | 0.99743+δ | 7.6E-4 | 0.49693 | 45 | 0.62878 | 0.37462-δ | 0.12311 |
| 14 | 0.49743+δ | 0.99924 | 0.99693 | 46 | 0.12878 | 0.62538-δ | 0.62311 |
| 15 | 0.49743+δ | 0.50076 | 0.49693 | 47 | 0.12878 | 0.87462-δ | 0.12311 |
| 16 | 0.99743+δ | 0.49924 | 0.99693 | 48 | 0.62878 | 0.12538-δ | 0.62311 |
| 17 | 0.74758-0.3*δ | 0.75639 | 0.2526 | 49 | 0.87599 | 0.62482+0.3*δ | 0.37671 |
| 18 | 0.24758-0.3*δ | 0.24361 | 0.7526 | 50 | 0.37599 | 0.37518+0.3*δ | 0.87671 |
| 19 | 0.24758-0.3*δ | 0.25639 | 0.2526 | 51 | 0.37599 | 0.12482+0.3*δ | 0.37671 |
| 20 | 0.74758-0.3*δ | 0.74361 | 0.7526 | 52 | 0.87599 | 0.87518+0.3*δ | 0.87671 |
| 21 | 0.75929-0.3*δ | 0.2522 | 0.25367 | 53 | 0.87543 | 0.13087+0.3*δ | 0.37437 |
| 22 | 0.25929-0.3*δ | 0.7478 | 0.75367 | 54 | 0.37543 | 0.86913+0.3*δ | 0.87437 |
| 23 | 0.25929-0.3*δ | 0.7522 | 0.25367 | 55 | 0.37543 | 0.63087+0.3*δ | 0.37437 |
| 24 | 0.75929-0.3*δ | 0.2478 | 0.75367 | 56 | 0.87543 | 0.36913+0.3*δ | 0.87437 |
| 25 | 0.00255-0.3*δ | 0.74372 | 0.75188 | 57 | 0.62566 | 0.62776-δ | 0.12584 |
| 26 | 0.50255-0.3*δ | 0.25628 | 0.25188 | 58 | 0.12566 | 0.37224-δ | 0.62584 |
| 27 | 0.50255-0.3*δ | 0.24372 | 0.75188 | 59 | 0.12566 | 0.12776-δ | 0.12584 |
| 28 | 0.00255-0.3*δ | 0.75628 | 0.25188 | 60 | 0.62566 | 0.87224-δ | 0.62584 |
| 29 | 0.00214-0.3*δ | 0.24588 | 0.75191 | 61 | 0.62788 | 0.12601-δ | 0.12652 |
| 30 | 0.50214-0.3*δ | 0.75412 | 0.25191 | 62 | 0.12788 | 0.87399-δ | 0.62652 |
| 31 | 0.50214-0.3*δ | 0.74588 | 0.75191 | 63 | 0.12788 | 0.62601-δ | 0.12652 |
| 32 | 0.00214-0.3*δ | 0.25412 | 0.25191 | 64 | 0.62788 | 0.37399-δ | 0.62652 |

**Table IV** $\Delta_5$-*x,y,z* type atomic displacement. $\delta_1$, $\delta_2$ are the atom deviations. $\delta_1$: $\delta_2$= 1

| Fe | x | y | z | Fe | x | y | z |
|---|---|---|---|---|---|---|---|
| 1 | 0.75057 | 0.99788+0.6*$\delta_1$ | 0.00227 | 33 | 0.87694 | 0.87945-$\delta_1$ | 0.37981+$\delta_2$ |
| 2 | 0.25057 | 0.00212-0.6*$\delta_1$ | 0.50227 | 34 | 0.37694 | 0.12055+$\delta_1$ | 0.87981+$\delta_2$ |
| 3 | 0.25057 | 0.49788+0.6*$\delta_1$ | 0.00227 | 35 | 0.37694 | 0.37945-$\delta_1$ | 0.37981-$\delta_2$ |
| 4 | 0.75057 | 0.50212-0.6*$\delta_1$ | 0.50212 | 36 | 0.87694 | 0.62055+$\delta_1$ | 0.87981-$\delta_2$ |
| 5 | 0.75116 | 0.49865+0.6*$\delta_1$ | 0.00111 | 37 | 0.87644 | 0.38747-$\delta_1$ | 0.38075-$\delta_2$ |
| 6 | 0.25116 | 0.50135-0.6*$\delta_1$ | 0.50111 | 38 | 0.37644 | 0.61253+$\delta_1$ | 0.88075-$\delta_2$ |
| 7 | 0.25116 | 0.99865+0.6*$\delta_1$ | 0.00111 | 39 | 0.37644 | 0.88747-$\delta_1$ | 0.38075+$\delta_2$ |
| 8 | 0.75116 | 0.00135-0.6*$\delta_1$ | 0.50111 | 40 | 0.87644 | 0.11253+$\delta_1$ | 0.88075+$\delta_2$ |
| 9 | 0.00187 | 0.5005-0.6*$\delta_1$ | 0.5017 | 41 | 0.62663 | 0.88662-0.6*$\delta_1$ | 0.12178+$\delta_2$ |
| 10 | 0.50187 | 0.4995+0.6*$\delta_1$ | 0.0017 | 42 | 0.12663 | 0.11338+0.6*$\delta_1$ | 0.62178+$\delta_2$ |
| 11 | 0.50187 | 5E-4-0.6*$\delta_1$ | 0.5017 | 43 | 0.12663 | 0.38662-0.6*$\delta_1$ | 0.12178-$\delta_2$ |
| 12 | 0.00187 | 0.9995+0.6*$\delta_1$ | 0.0017 | 44 | 0.62663 | 0.61338+0.6*$\delta_1$ | 0.62178-$\delta_2$ |
| 13 | 0.99743 | 7.6E-4-0.6*$\delta_1$ | 0.49693 | 45 | 0.62878 | 0.37462-0.6*$\delta_1$ | 0.12311-$\delta_2$ |
| 14 | 0.49743 | 0.99924+0.6*$\delta_1$ | 0.99693 | 46 | 0.12878 | 0.62538+0.6*$\delta_1$ | 0.62311-$\delta_2$ |
| 15 | 0.49743 | 0.50076-0.6*$\delta_1$ | 0.49693 | 47 | 0.12878 | 0.87462-0.6*$\delta_1$ | 0.12311+$\delta_2$ |
| 16 | 0.99743 | 0.49924+0.6*$\delta_1$ | 0.99693 | 48 | 0.62878 | 0.12538+0.6*$\delta_1$ | 0.62311+$\delta_2$ |
| 17 | 0.74758 | 0.75639-$\delta_1$ | 0.2526 | 49 | 0.87599 | 0.62482-$\delta_1$ | 0.37671-$\delta_2$ |
| 18 | 0.24758 | 0.24361+$\delta_1$ | 0.7526 | 50 | 0.37599 | 0.37518+$\delta_1$ | 0.87671-$\delta_2$ |
| 19 | 0.24758 | 0.25639-$\delta_1$ | 0.2526 | 51 | 0.37599 | 0.12482-$\delta_1$ | 0.37671+$\delta_2$ |
| 20 | 0.74758 | 0.74361+$\delta_1$ | 0.7526 | 52 | 0.87599 | 0.87518+$\delta_1$ | 0.87671+$\delta_2$ |
| 21 | 0.75929 | 0.2522-$\delta_1$ | 0.25367 | 53 | 0.87543 | 0.13087-$\delta_1$ | 0.37437+$\delta_2$ |
| 22 | 0.25929 | 0.7478+$\delta_1$ | 0.75367 | 54 | 0.37543 | 0.86913+$\delta_1$ | 0.87437+$\delta_2$ |
| 23 | 0.25929 | 0.7522-$\delta_1$ | 0.25367 | 55 | 0.37543 | 0.63087-$\delta_1$ | 0.37437-$\delta_2$ |
| 24 | 0.75929 | 0.2478+$\delta_1$ | 0.75367 | 56 | 0.87543 | 0.36913+$\delta_1$ | 0.87437-$\delta_2$ |
| 25 | 0.00255 | 0.74372+$\delta_1$ | 0.75188 | 57 | 0.62566 | 0.62776-0.6*$\delta_1$ | 0.12584-$\delta_2$ |
| 26 | 0.50255 | 0.25628-$\delta_1$ | 0.25188 | 58 | 0.12566 | 0.37224+0.6*$\delta_1$ | 0.62584-$\delta_2$ |
| 27 | 0.50255 | 0.24372+$\delta_1$ | 0.75188 | 59 | 0.12566 | 0.12776-0.6*$\delta_1$ | 0.12584+$\delta_2$ |
| 28 | 0.00255 | 0.75628-$\delta_1$ | 0.25188 | 60 | 0.62566 | 0.87224+0.6*$\delta_1$ | 0.62584+$\delta_2$ |
| 29 | 0.00214 | 0.24588+$\delta_1$ | 0.75191 | 61 | 0.62788 | 0.12601-0.6*$\delta_1$ | 0.12652+$\delta_2$ |
| 30 | 0.50214 | 0.75412-$\delta_1$ | 0.25191 | 62 | 0.12788 | 0.87399+0.6*$\delta_1$ | 0.62652+$\delta_2$ |
| 31 | 0.50214 | 0.74588+$\delta_1$ | 0.75191 | 63 | 0.12788 | 0.62601-0.6*$\delta_1$ | 0.12652-$\delta_2$ |
| 32 | 0.00214 | 0.25412-$\delta_1$ | 0.25191 | 64 | 0.62788 | 0.37399+0.6*$\delta_1$ | 0.62652-$\delta_2$ |

**Table V** $\Delta_4$ type atomic displacement. $\delta_1$, $\delta_2$ are the atom deviations.

| Fe | x | y | z | Fe | x | y | z |
|---|---|---|---|---|---|---|---|
| 1 | 0.75057-$\delta_1$ | 0.99788 | 0.00227-$\delta_2$ | 33 | 0.87694 | 0.87945+$\delta_1$ | 0.37981+$\delta_2$ |
| 2 | 0.25057+$\delta_1$ | 0.00212 | 0.50227-$\delta_2$ | 34 | 0.37694 | 0.12055+$\delta_1$ | 0.87981+$\delta_2$ |
| 3 | 0.25057-$\delta_1$ | 0.49788 | 0.00227-$\delta_2$ | 35 | 0.37694 | 0.37945+$\delta_1$ | 0.37981+$\delta_2$ |
| 4 | 0.75057+$\delta_1$ | 0.50212 | 0.50212-$\delta_2$ | 36 | 0.87694 | 0.62055+$\delta_1$ | 0.87981+$\delta_2$ |
| 5 | 0.75116-$\delta_1$ | 0.49865 | 0.00111-$\delta_2$ | 37 | 0.87644 | 0.38747+$\delta_1$ | 0.38075+$\delta_2$ |
| 6 | 0.25116+$\delta_1$ | 0.50135 | 0.50111-$\delta_2$ | 38 | 0.37644 | 0.61253+$\delta_1$ | 0.88075+$\delta_2$ |
| 7 | 0.25116-$\delta_1$ | 0.99865 | 0.00111-$\delta_2$ | 39 | 0.37644 | 0.88747+$\delta_1$ | 0.38075+$\delta_2$ |
| 8 | 0.75116+$\delta_1$ | 0.00135 | 0.50111-$\delta_2$ | 40 | 0.87644 | 0.11253+$\delta_1$ | 0.88075+$\delta_2$ |
| 9 | 0.00187-$\delta_1$ | 0.5005 | 0.5017-$\delta_2$ | 41 | 0.62663 | 0.88662+$\delta_1$ | 0.12178+$\delta_2$ |
| 10 | 0.50187+$\delta_1$ | 0.4995 | 0.0017-$\delta_2$ | 42 | 0.12663 | 0.11338+$\delta_1$ | 0.62178+$\delta_2$ |
| 11 | 0.50187-$\delta_1$ | 5E-4 | 0.5017-$\delta_2$ | 43 | 0.12663 | 0.38662+$\delta_1$ | 0.12178+$\delta_2$ |
| 12 | 0.00187+$\delta_1$ | 0.9995 | 0.0017-$\delta_2$ | 44 | 0.62663 | 0.61338+$\delta_1$ | 0.62178+$\delta_2$ |
| 13 | 0.99743-$\delta_1$ | 7.6E-4 | 0.49693-$\delta_2$ | 45 | 0.62878 | 0.37462+$\delta_1$ | 0.12311+$\delta_2$ |
| 14 | 0.49743+$\delta_1$ | 0.99924 | 0.99693-$\delta_2$ | 46 | 0.12878 | 0.62538+$\delta_1$ | 0.62311+$\delta_2$ |
| 15 | 0.49743-$\delta_1$ | 0.50076 | 0.49693-$\delta_2$ | 47 | 0.12878 | 0.87462+$\delta_1$ | 0.12311+$\delta_2$ |
| 16 | 0.99743+$\delta_1$ | 0.49924 | 0.99693-$\delta_2$ | 48 | 0.62878 | 0.12538+$\delta_1$ | 0.62311+$\delta_2$ |
| 17 | 0.74758-$\delta_1$ | 0.75639 | 0.2526-$\delta_2$ | 49 | 0.87599 | 0.62482-$\delta_1$ | 0.37671+$\delta_2$ |
| 18 | 0.24758+$\delta_1$ | 0.24361 | 0.7526-$\delta_2$ | 50 | 0.37599 | 0.37518-$\delta_1$ | 0.87671+$\delta_2$ |
| 19 | 0.24758-$\delta_1$ | 0.25639 | 0.2526-$\delta_2$ | 51 | 0.37599 | 0.12482-$\delta_1$ | 0.37671+$\delta_2$ |
| 20 | 0.74758+$\delta_1$ | 0.74361 | 0.7526-$\delta_2$ | 52 | 0.87599 | 0.87518-$\delta_1$ | 0.87671+$\delta_2$ |
| 21 | 0.75929-$\delta_1$ | 0.2522 | 0.25367-$\delta_2$ | 53 | 0.87543 | 0.13087-$\delta_1$ | 0.37437+$\delta_2$ |
| 22 | 0.25929+$\delta_1$ | 0.7478 | 0.75367-$\delta_2$ | 54 | 0.37543 | 0.86913-$\delta_1$ | 0.87437+$\delta_2$ |
| 23 | 0.25929-$\delta_1$ | 0.7522 | 0.25367-$\delta_2$ | 55 | 0.37543 | 0.63087-$\delta_1$ | 0.37437+$\delta_2$ |
| 24 | 0.75929+$\delta_1$ | 0.2478 | 0.75367-$\delta_2$ | 56 | 0.87543 | 0.36913-$\delta_1$ | 0.87437+$\delta_2$ |
| 25 | 0.00255-$\delta_1$ | 0.74372 | 0.75188-$\delta_2$ | 57 | 0.62566 | 0.62776-$\delta_1$ | 0.12584+$\delta_2$ |
| 26 | 0.50255+$\delta_1$ | 0.25628 | 0.25188-$\delta_2$ | 58 | 0.12566 | 0.37224-$\delta_1$ | 0.62584+$\delta_2$ |
| 27 | 0.50255-$\delta_1$ | 0.24372 | 0.75188-$\delta_2$ | 59 | 0.12566 | 0.12776-$\delta_1$ | 0.12584+$\delta_2$ |
| 28 | 0.00255+$\delta_1$ | 0.75628 | 0.25188-$\delta_2$ | 60 | 0.62566 | 0.87224-$\delta_1$ | 0.62584+$\delta_2$ |
| 29 | 0.00214-$\delta_1$ | 0.24588 | 0.75191-$\delta_2$ | 61 | 0.62788 | 0.12601-$\delta_1$ | 0.12652+$\delta_2$ |
| 30 | 0.50214+$\delta_1$ | 0.75412 | 0.25191-$\delta_2$ | 62 | 0.12788 | 0.87399-$\delta_1$ | 0.62652+$\delta_2$ |
| 31 | 0.50214-$\delta_1$ | 0.74588 | 0.75191-$\delta_2$ | 63 | 0.12788 | 0.62601-$\delta_1$ | 0.12652+$\delta_2$ |
| 32 | 0.00214+$\delta_1$ | 0.25412 | 0.25191-$\delta_2$ | 64 | 0.62788 | 0.37399-$\delta_1$ | 0.62652+$\delta_2$ |

Table VI $W_1$ type atomic displacement. $\delta$ is the atom deviation.

| Fe | x | y | z | Fe | x | y | z |
|---|---|---|---|---|---|---|---|
| 1 | 0.75057+δ | 0.99788 | 0.00227+0.5*δ | 33 | 0.87694 | 0.87945 | 0.37981 |
| 2 | 0.25057-δ | 0.00212 | 0.50227-0.5*δ | 34 | 0.37694 | 0.12055 | 0.87981 |
| 3 | 0.25057-δ | 0.49788 | 0.00227-0.5*δ | 35 | 0.37694 | 0.37945 | 0.37981 |
| 4 | 0.75057+δ | 0.50212 | 0.50212+0.5*δ | 36 | 0.87694 | 0.62055 | 0.87981 |
| 5 | 0.75116-δ | 0.49865 | 0.00111-0.5*δ | 37 | 0.87644 | 0.38747 | 0.38075 |
| 6 | 0.25116+δ | 0.50135 | 0.50111+0.5*δ | 38 | 0.37644 | 0.61253 | 0.88075 |
| 7 | 0.25116+δ | 0.99865 | 0.00111+0.5*δ | 39 | 0.37644 | 0.88747 | 0.38075 |
| 8 | 0.75116-δ | 0.00135 | 0.50111-0.5*δ | 40 | 0.87644 | 0.11253 | 0.88075 |
| 9 | 0.00187+δ | 0.5005 | 0.5017-0.5*δ | 41 | 0.62663 | 0.88662+δ | 0.12178-0.5*δ |
| 10 | 0.50187-δ | 0.4995 | 0.0017+0.5*δ | 42 | 0.12663 | 0.11338-δ | 0.62178-0.5*δ |
| 11 | 0.50187-δ | 5E-4 | 0.5017+0.5*δ | 43 | 0.12663 | 0.38662+δ | 0.12178-0.5*δ |
| 12 | 0.00187+δ | 0.9995 | 0.0017-0.5*δ | 44 | 0.62663 | 0.61338-δ | 0.62178-0.5*δ |
| 13 | 0.99743-δ | 7.6E-4 | 0.49693+0.5*δ | 45 | 0.62878 | 0.37462-δ | 0.12311+0.5*δ |
| 14 | 0.49743+δ | 0.99924 | 0.99693-0.5*δ | 46 | 0.12878 | 0.62538+δ | 0.62311+0.5*δ |
| 15 | 0.49743+δ | 0.50076 | 0.49693-0.5*δ | 47 | 0.12878 | 0.87462-δ | 0.12311+0.5*δ |
| 16 | 0.99743-δ | 0.49924 | 0.99693+0.5*δ | 48 | 0.62878 | 0.12538+δ | 0.62311+0.5*δ |
| 17 | 0.74758 | 0.75639 | 0.2526 | 49 | 0.87599 | 0.62482 | 0.37671 |
| 18 | 0.24758 | 0.24361 | 0.7526 | 50 | 0.37599 | 0.37518 | 0.87671 |
| 19 | 0.24758 | 0.25639 | 0.2526 | 51 | 0.37599 | 0.12482 | 0.37671 |
| 20 | 0.74758 | 0.74361 | 0.7526 | 52 | 0.87599 | 0.87518 | 0.87671 |
| 21 | 0.75929 | 0.2522 | 0.25367 | 53 | 0.87543 | 0.13087 | 0.37437 |
| 22 | 0.25929 | 0.7478 | 0.75367 | 54 | 0.37543 | 0.86913 | 0.87437 |
| 23 | 0.25929 | 0.7522 | 0.25367 | 55 | 0.37543 | 0.63087 | 0.37437 |
| 24 | 0.75929 | 0.2478 | 0.75367 | 56 | 0.87543 | 0.36913 | 0.87437 |
| 25 | 0.00255 | 0.74372 | 0.75188 | 57 | 0.62566 | 0.62776+δ | 0.12584+0.5*δ |
| 26 | 0.50255 | 0.25628 | 0.25188 | 58 | 0.12566 | 0.37224-δ | 0.62584+0.5*δ |
| 27 | 0.50255 | 0.24372 | 0.75188 | 59 | 0.12566 | 0.12776+δ | 0.12584+0.5*δ |
| 28 | 0.00255 | 0.75628 | 0.25188 | 60 | 0.62566 | 0.87224-δ | 0.62584+0.5*δ |
| 29 | 0.00214 | 0.24588 | 0.75191 | 61 | 0.62788 | 0.12601-δ | 0.12652-0.5*δ |
| 30 | 0.50214 | 0.75412 | 0.25191 | 62 | 0.12788 | 0.87399+δ | 0.62652-0.5*δ |
| 31 | 0.50214 | 0.74588 | 0.75191 | 63 | 0.12788 | 0.62601-δ | 0.12652-0.5*δ |
| 32 | 0.00214 | 0.25412 | 0.25191 | 64 | 0.62788 | 0.37399+δ | 0.62652-0.5*δ |

Table VII $W_2$ type atomic displacement. δ is the atom deviation.

| Fe | x | y | z | Fe | x | y | z |
|---|---|---|---|---|---|---|---|
| 1 | 0.75057-δ | 0.99788 | 0.00227-0.5*δ | 33 | 0.87694 | 0.87945 | 0.37981 |
| 2 | 0.25057+δ | 0.00212 | 0.50227+0.5*δ | 34 | 0.37694 | 0.12055 | 0.87981 |
| 3 | 0.25057+δ | 0.49788 | 0.00227+0.5*δ | 35 | 0.37694 | 0.37945 | 0.37981 |
| 4 | 0.75057-δ | 0.50212 | 0.50212-0.5*δ | 36 | 0.87694 | 0.62055 | 0.87981 |
| 5 | 0.75116+δ | 0.49865 | 0.00111+0.5*δ | 37 | 0.87644 | 0.38747 | 0.38075 |
| 6 | 0.25116-δ | 0.50135 | 0.50111-0.5*δ | 38 | 0.37644 | 0.61253 | 0.88075 |
| 7 | 0.25116-δ | 0.99865 | 0.00111-0.5*δ | 39 | 0.37644 | 0.88747 | 0.38075 |
| 8 | 0.75116+δ | 0.00135 | 0.50111+0.5*δ | 40 | 0.87644 | 0.11253 | 0.88075 |
| 9 | 0.00187-δ | 0.5005 | 0.5017+0.5*δ | 41 | 0.62663 | 0.88662+δ | 0.12178-0.5*δ |
| 10 | 0.50187+δ | 0.4995 | 0.0017-0.5*δ | 42 | 0.12663 | 0.11338-δ | 0.62178-0.5*δ |
| 11 | 0.50187+δ | 5E-4 | 0.5017-0.5*δ | 43 | 0.12663 | 0.38662+δ | 0.12178-0.5*δ |
| 12 | 0.00187-δ | 0.9995 | 0.0017+0.5*δ | 44 | 0.62663 | 0.61338-δ | 0.62178-0.5*δ |
| 13 | 0.99743+δ | 7.6E-4 | 0.49693-0.5*δ | 45 | 0.62878 | 0.37462-δ | 0.12311+0.5*δ |
| 14 | 0.49743-δ | 0.99924 | 0.99693+0.5*δ | 46 | 0.12878 | 0.62538+δ | 0.62311+0.5*δ |
| 15 | 0.49743-δ | 0.50076 | 0.49693+0.5*δ | 47 | 0.12878 | 0.87462-δ | 0.12311+0.5*δ |
| 16 | 0.99743+δ | 0.49924 | 0.99693-0.5*δ | 48 | 0.62878 | 0.12538+δ | 0.62311+0.5*δ |
| 17 | 0.74758 | 0.75639 | 0.2526 | 49 | 0.87599 | 0.62482 | 0.37671 |
| 18 | 0.24758 | 0.24361 | 0.7526 | 50 | 0.37599 | 0.37518 | 0.87671 |
| 19 | 0.24758 | 0.25639 | 0.2526 | 51 | 0.37599 | 0.12482 | 0.37671 |
| 20 | 0.74758 | 0.74361 | 0.7526 | 52 | 0.87599 | 0.87518 | 0.87671 |
| 21 | 0.75929 | 0.2522 | 0.25367 | 53 | 0.87543 | 0.13087 | 0.37437 |
| 22 | 0.25929 | 0.7478 | 0.75367 | 54 | 0.37543 | 0.86913 | 0.87437 |
| 23 | 0.25929 | 0.7522 | 0.25367 | 55 | 0.37543 | 0.63087 | 0.37437 |
| 24 | 0.75929 | 0.2478 | 0.75367 | 56 | 0.87543 | 0.36913 | 0.87437 |
| 25 | 0.00255 | 0.74372 | 0.75188 | 57 | 0.62566 | 0.62776+δ | 0.12584+0.5*δ |
| 26 | 0.50255 | 0.25628 | 0.25188 | 58 | 0.12566 | 0.37224-δ | 0.62584+0.5*δ |
| 27 | 0.50255 | 0.24372 | 0.75188 | 59 | 0.12566 | 0.12776+δ | 0.12584+0.5*δ |
| 28 | 0.00255 | 0.75628 | 0.25188 | 60 | 0.62566 | 0.87224-δ | 0.62584+0.5*δ |
| 29 | 0.00214 | 0.24588 | 0.75191 | 61 | 0.62788 | 0.12601-δ | 0.12652-0.5*δ |
| 30 | 0.50214 | 0.75412 | 0.25191 | 62 | 0.12788 | 0.87399+δ | 0.62652-0.5*δ |
| 31 | 0.50214 | 0.74588 | 0.75191 | 63 | 0.12788 | 0.62601-δ | 0.12652-0.5*δ |
| 32 | 0.00214 | 0.25412 | 0.25191 | 64 | 0.62788 | 0.37399+δ | 0.62652-0.5*δ |

## Figures

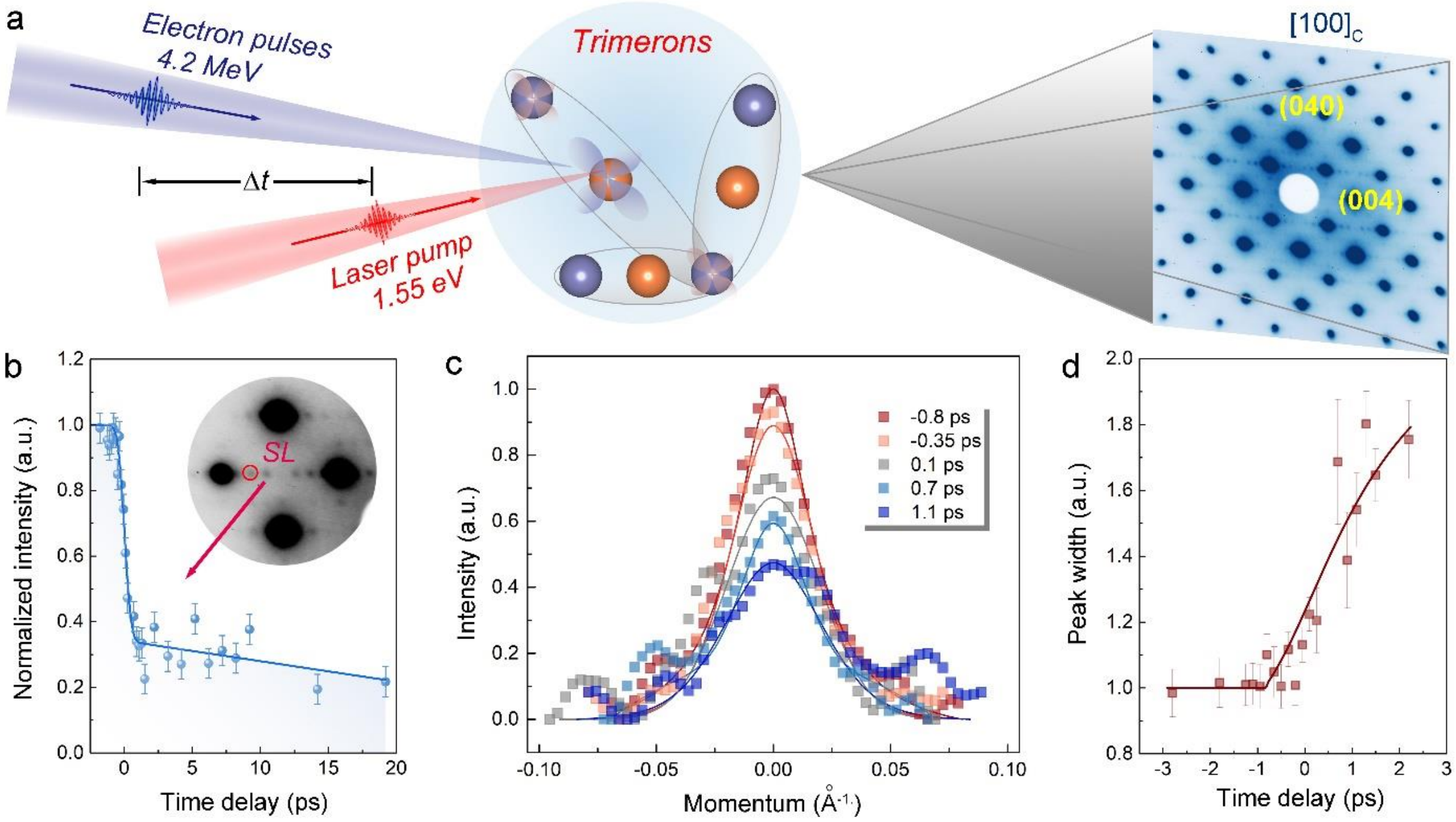

**FIG. 1.** UED experiment schematic diagram and SL reflections variation upon photoexcitation. **a** $Fe_3O_4$ single crystal is photoexcited with ultrashort laser pulses with 1.55 eV and is probed by ultrashort electron pulses with 4.2 MeV. The UED pattern of the monoclinic phase with trimeron order at 34 K along the $[100]_{cubic}$ orientation is shown on the right. **b** A representative integrated intensity variation as a function of delay time, averaged over six SL reflections, $(0, \bar{2}, 2+\frac{1}{2})$, $(0, \bar{2}, 3)$, $(0, \bar{2}, 3+\frac{1}{2})$, $(0, \bar{2}, 4+\frac{1}{2})$, $(0, \bar{2}, 5)$, $(0, \bar{2}, 5+\frac{1}{2})$, between $(0, \bar{2}, 2)$ and $(0, \bar{2}, 6)$ Bragg peaks in the insert. The solid line is a guide to the eye. Error bars represent the standard deviation in the mean of intensity before time zero. The insert shows four Bragg peaks $(0, \bar{2}, 2)$, $(0, \bar{4}, 4)$, $(0, \bar{2}, 6)$, $(0, 0, 4)$ with surrounding SL reflections **c** Line profile of $(0, \bar{2}, 3)$ SL reflection at early time delays. The SL position is labeled by a red circle in **b**. The experiment data are shown as squares symbols and the solid lines are the fitted results. **d** Peak width (FWHM) measurements from one SL reflection at short time delays.

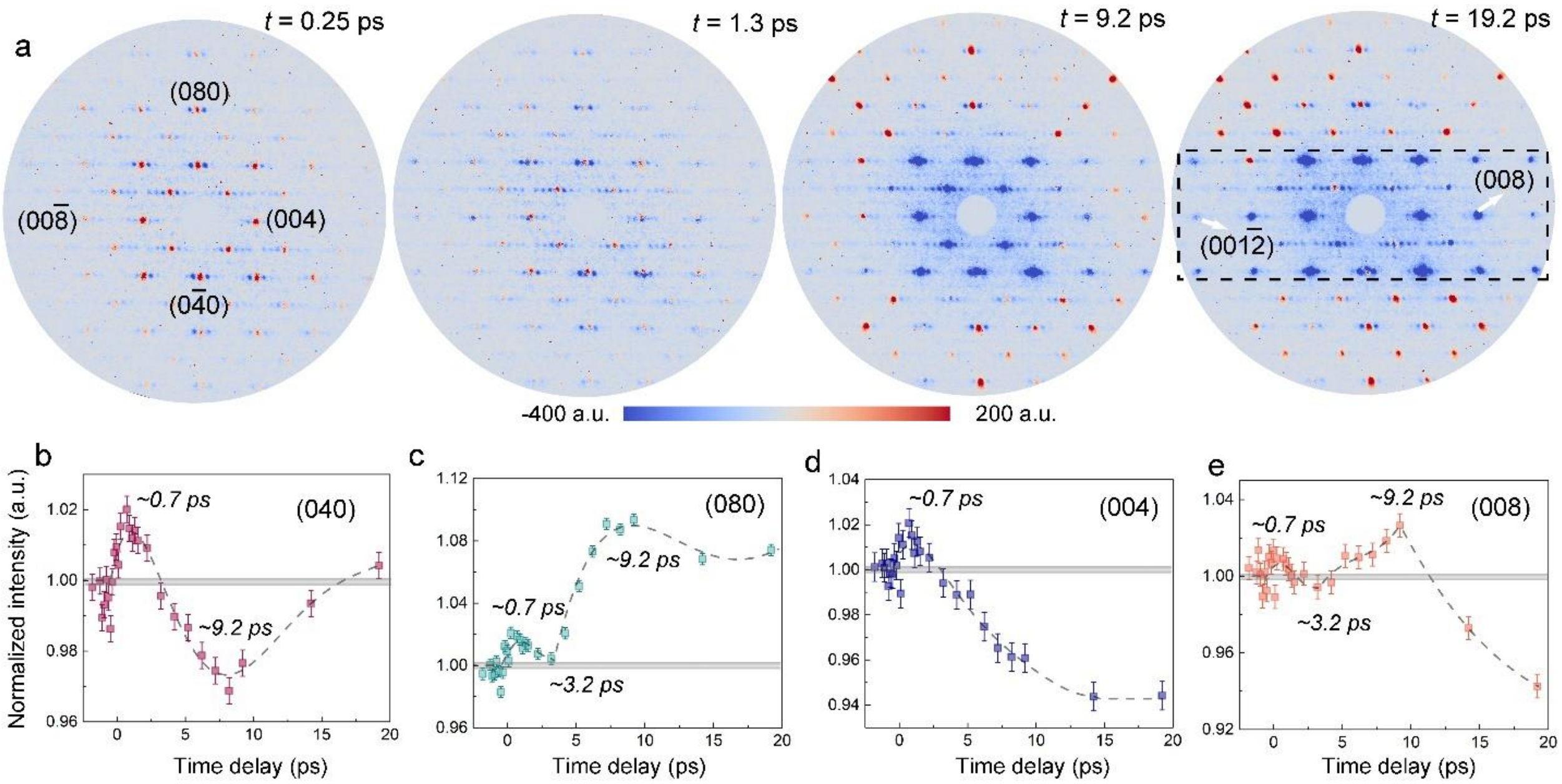

**FIG. 2.** Intensity variation measurement at 5 mJ·cm$^{-2}$. **a** Intensity difference map $\Delta I(\boldsymbol{q}, t) = I(\boldsymbol{q}, t) - I(\boldsymbol{q}, t_0)$ for a few representative time delays at pump fluence 5 mJ·cm$^{-2}$, where $t_0$ represents the time before the pump arrives and $\boldsymbol{q}$ is the scattering vector in the reciprocal space The negative change (in blue color) indicates the intensity decreases after time zero, the positive change (in red color) indicates the intensity increases after time zero. The dashed box in the pattern at 19.2 ps highlights the main changes, compared with the pattern at 9.2 ps. **b**-**e** Intensity as a function of time measured from *four* Bragg peaks. The time delays values, 0.7 ps, 3.2 ps and 9.2 ps, are labeled in each plot. Error bars represent the standard deviation in the mean of intensity before time zero. The peak positions of *b*-(040), *c*-(080), *d*-(004), and *e*-(008) are labeled in **a**.

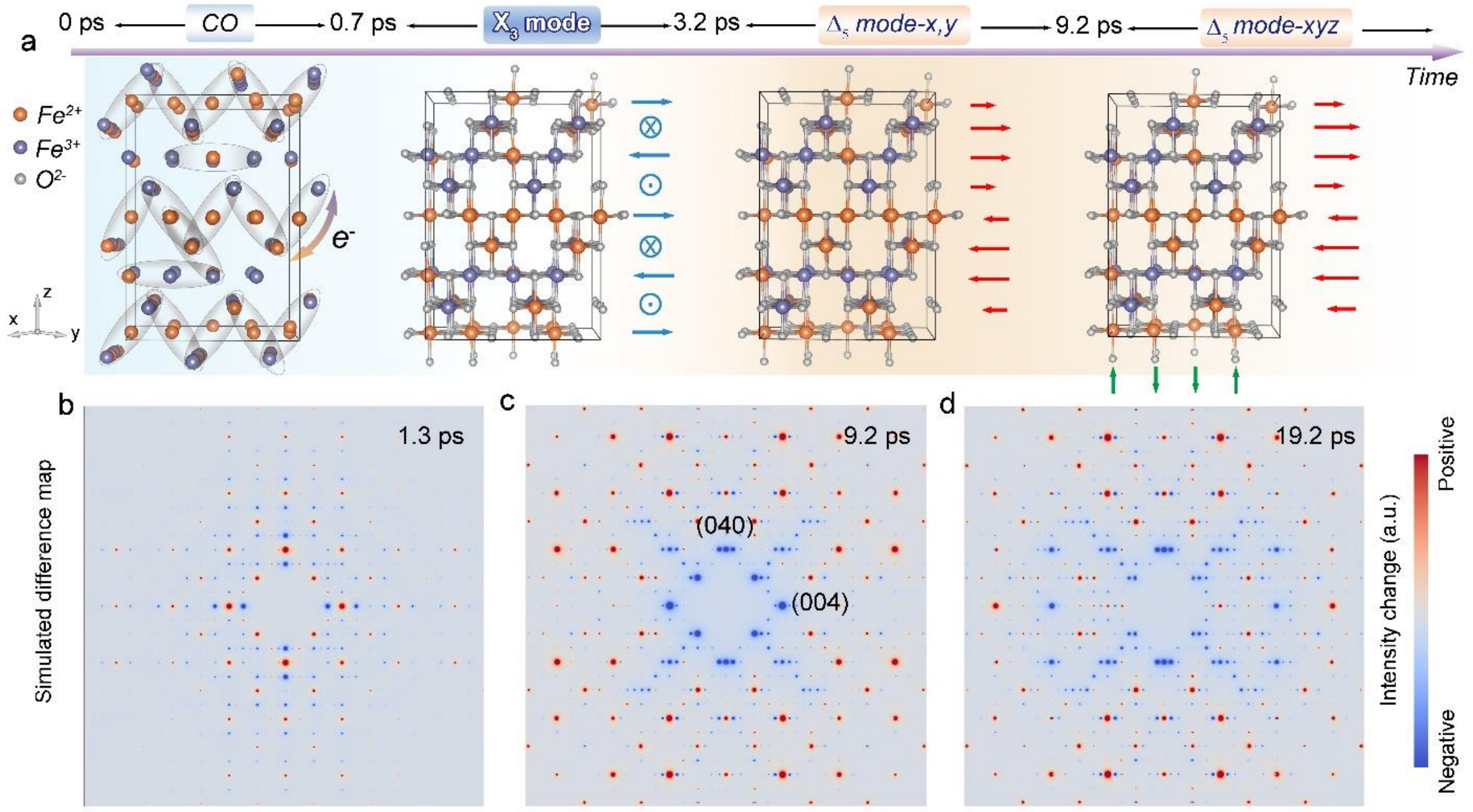

**FIG. 3.** Schematic diagram of dynamic behaviors of the lattice following photoexcitation and the corresponding diffraction simulation results at 5 mJ·cm$^{-2}$. **a** In the first 0.7 ps, the charge discrepancy between Fe ions in the trimerons is reduced due to the excitation of charge transfer between Fe ions. A projected unit cell in the low-temperature monoclinic phase along [110] direction is shown. The trimerons are highlighted by translucent ellipsoids. Fe ions on the tetrahedral sites are omitted for clarity. Then the $X_3$ phonon modes are excited after 0.7 ps via the energy flow from electrons to lattice. The blue arrows indicate the displacement of Fe ions in the sample layer, along the ***-x+y*** and ***x-y*** directions. The circles marked "·" and "×" show the displacement of Fe ions is along the ***x+y*** and ***-x-y*** directions. After ~ 3.2 ps, $\Delta_5$ phonon modes are excited via phonon-phonon interactions. Between 3.2-9.2 ps, the atomic displacements mainly happen in the ***x-y*** plane as shown in the red arrows. After 9.2 ps, the atomic displacement along the ***z*** direction dominates based on the ***x-y*** plane distortion. The red arrow on the right side indicates the ***x-y*** displacement of all the Fe ions in each layer, and the length indicates the displacement amplitude. The green arrows at the bottom show the displacement along ***z*** direction of Fe in each Fe-O layer. The atomic displacement pattern after 9.2 ps is the combination of three-dimensional (***x***, ***y***, ***z***) displacement. Simulated intensity difference patterns at the corresponding time delays, 1.3 ps, 9.2 ps and 19.2 ps are shown in **b**-**d**. Simulation parameter in **b** includes the reduced charge discrepancy and lattice displacement with the $X_3$ mode; simulation in **c** is the atomic displacement corresponding to the $\Delta_5$ *mode-x, y*, based on the reduced charge ordering state and the $X_3$-type displacements in **b**; simulated model in **d** is the atomic displacement along ***z*** direction based on the distorted structure in **c**, i.e., $\Delta_5$ *mode-x, y, z*. The simulation results are qualitatively consistent with experimental observations.

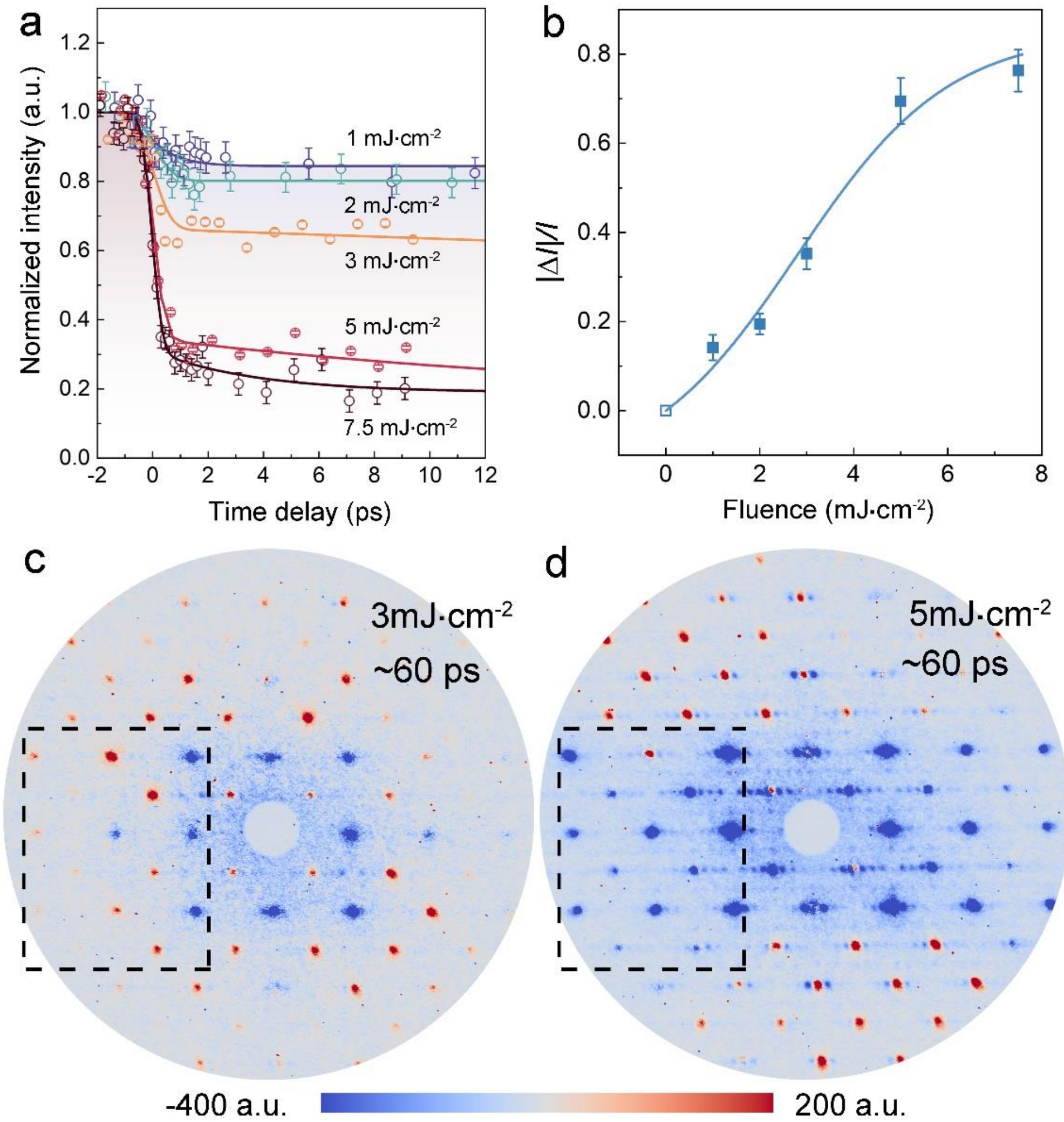

**FIG. 4.** Pump-fluence dependent dynamic behaviors. **a** The SL reflection intensity variation with time delays at different pump fluences. **b** The maximum intensity variation ($|\Delta I|$) at 9.2 ps extracted from different pump fluences in **a**. The curve is the fitted result for a guide to the eye. The open box in the plot shows the (0, 0) point. **c**, **d** Intensity difference map at ~60 ps with the pump fluence of 3 mJ·cm$^{-2}$ and 5 mJ·cm$^{-2}$, respectively. At 3 mJ·cm$^{-2}$, the intensity variation is smaller than that at 5 mJ·cm$^{-2}$. The significant difference between **c** and **d** is the intensities of the reflections in the frames, which exhibit a different tendency at ~60 ps, indicating the different lattice displacement on the long timescale at 3 and 5 mJ·cm$^{-2}$.

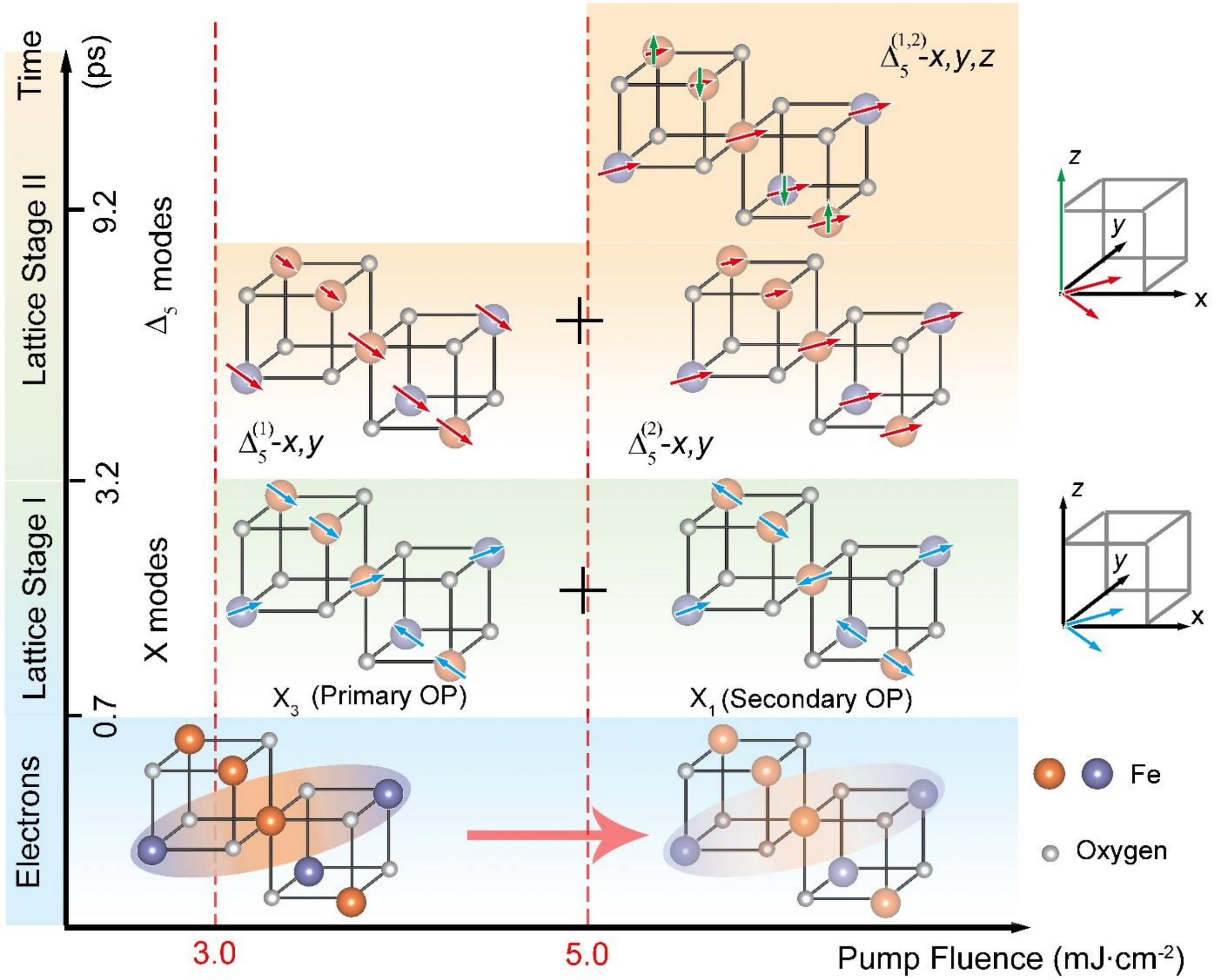

**FIG. 5.** Photoinduced dynamic processes as a function of time delay and laser pump fluence in $Fe_3O_4$. After photoexcitation with 1.55 eV laser pulses, the electronic state is excited and charge discrepancy in trimerons has been reduced at early time delays (0-0.7 ps). The Fe-O cubes present a partial crystal structure. Above 3 mJ·cm$^{-2}$, the lattice distortion is triggered and follow different types of phonon modes on two timescales after 0.7 ps. In Stage I (0.7 ps-3.2 ps), the electrons drive the lattice distortions with $X_3$- and $X_1$-type phonon modes, which are the primary and secondary order parameter (OP) for the Verwey transition, respectively. The blue arrows indicate the ***x*-*y*** plane displacements along the diagonal directions in the cube. After ~3.2 ps, the ***x*-*y*** plane lattice distortions with $\Delta_5$-type modes ($\Delta_5$-*x*, *y*) become dominant, leading the system to Stage II. $\Delta_5^{(1)}$-***x***, ***y*** and $\Delta_5^{(2)}$-***x***, ***y*** modes are two degenerate phonon modes. The red arrows indicate the ***x*-*y*** plane displacements. Above 5 mJ·cm$^{-2}$, an additional atomic displacement along ***z*** in $\Delta_5$ modes ($\Delta_5$-*x*, *y*, *z*) is observed after 9.2 ps, as shown by green arrows.

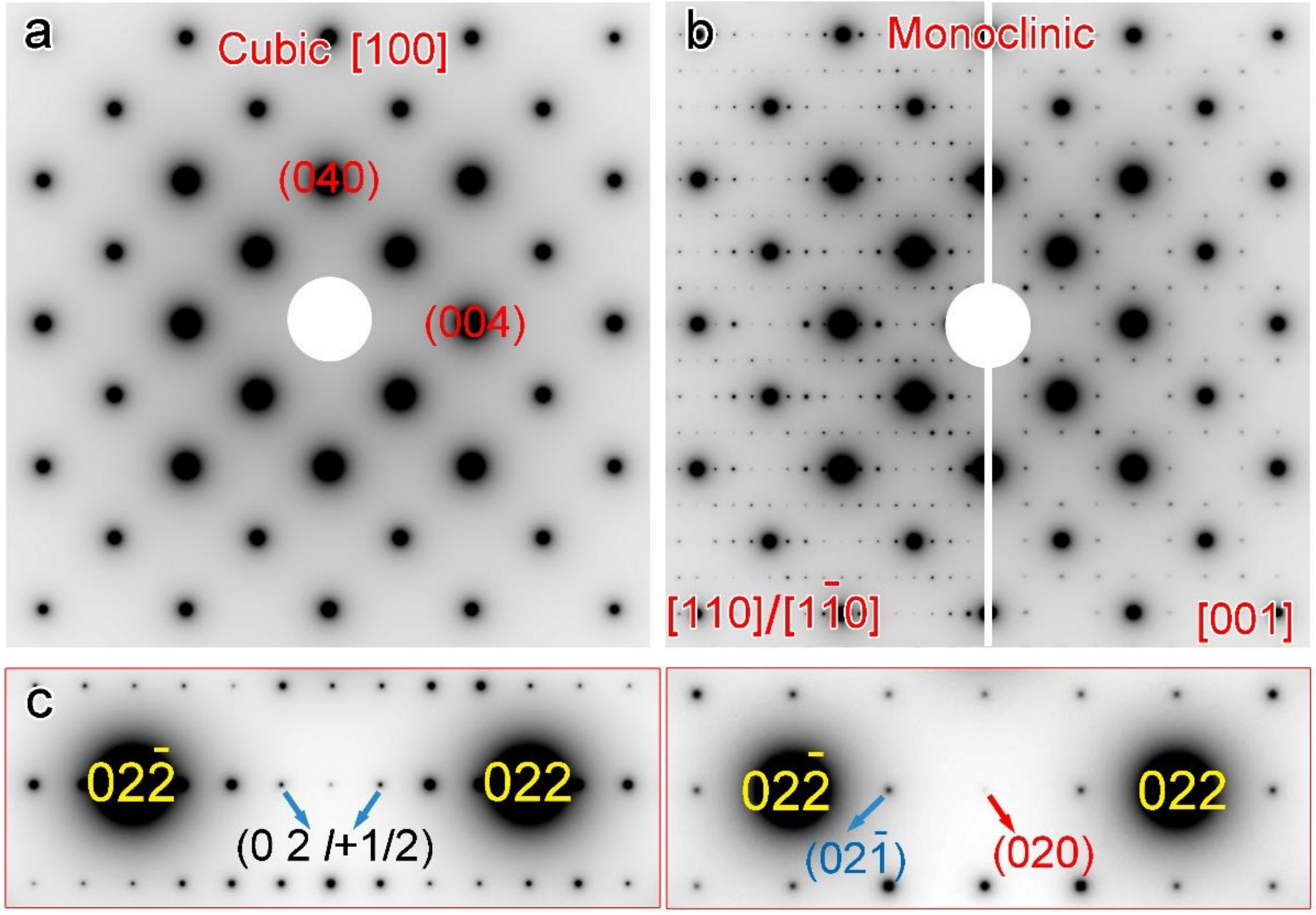

**FIG. 6.** Electron diffraction pattern simulation results for high-temperature phase and low-temperature phase. **a** Simulated diffraction pattern along [100] direction for the high-temperature cubic structure. **b** Left panel: simulated diffraction pattern along $[110]_{monoclinic}$, and $[1\bar{1}0]_{monoclinic}$ directions for the low-temperature monoclinic structure; right panel: simulated diffraction pattern along $[001]_{monoclinic}$. **c** Part of Bragg peaks and SL reflections shown in **b**, showing $(0, k, l+1/2)_{cubic}$ and $(0, k, l)_{cubic}$ types of SL reflections. If the [100] direction in the cubic structure is transferred into [110]and/or $[1\bar{1}0]$ in the monoclinic phase, the $(0, k, l+1/2)_{cubic}$ will be observed. If the [100] direction in the cubic structure is transferred into [001] in the monoclinic phase, only the $(0, k, l)_{cubic}$ will be observed in the diffraction pattern.

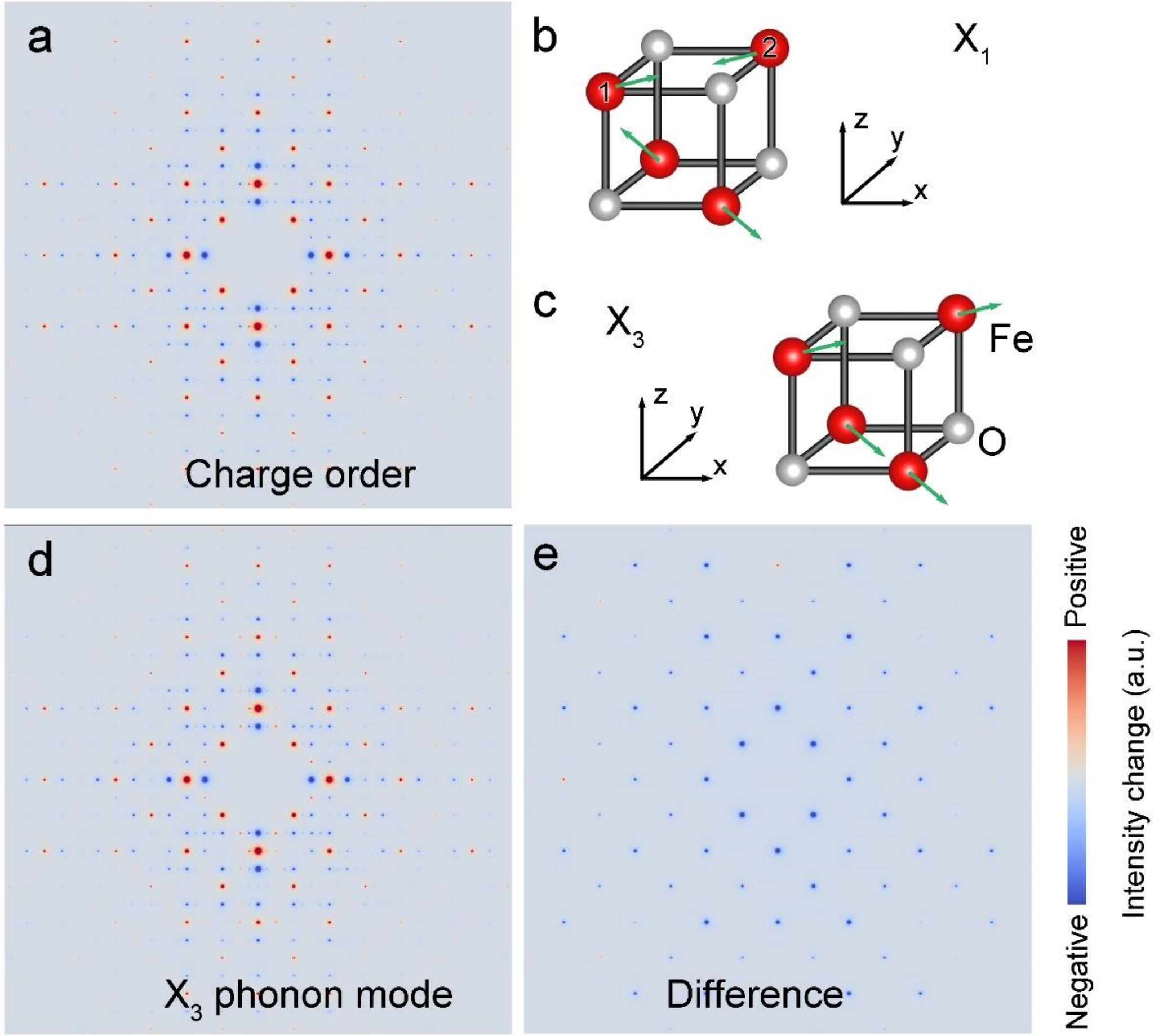

**FIG. 7.** Simulated intensity difference patterns. **a** Intensity difference map at 0.7 ps induced by the melting of charge ordering state. **b**, **c** atomic displacement in $X_1$ and $X_3$ modes. The arrows indicate the displacement direction. The red spheres represent Fe ions in the octahedral sites and the grey spheres are the oxygens. **d** Intensity difference map for one $X_3$ phonon mode. **e** Intensity difference between **a** and **d**, i.e., *I*(**e**) = *I*(**d**) – *I*(**a**), indicating the intensity change induced by lattice distortion following the $X_3$ phonon mode.